\documentclass[twocolumn]{aastex701}

\usepackage{mathptmx}
\usepackage[T1]{fontenc}
\usepackage{ae,aecompl}
\usepackage{graphicx}	
\graphicspath{{./Figures/}} 
\usepackage{amsmath}	
\usepackage{amssymb}	
\usepackage{subfigmat}
\usepackage{enumerate}
\usepackage{multirow}
\usepackage{natbib}
\usepackage{tabularx}
\usepackage{array}
\usepackage{caption}
\usepackage{longtable}

\newcommand{\kms}{\hbox{\;km s$^{-1}$}}
\newcommand{\ms}{\hbox{\;m s$^{-1}$}}

\usepackage{lineno}
\usepackage{tabularx, booktabs}
\usepackage{adjustbox}
\usepackage{longtable}
\usepackage{ltablex}
\usepackage{threeparttablex}
\usepackage{amssymb}
\graphicspath{{./}{figures/}}

\def\gtrsim{\mathrel{\hbox{\rlap{\hbox{\lower5pt\hbox{$\sim$}}}\hbox{$>$}}}}
\newcommand{\au}{\;\mbox{au}}
\newcommand{\kyr}{\;\mbox{kyr}}
\newcommand{\msun}{\;\mbox{M$_\odot$}}
\newcommand{\mdotyr}{\hbox{\msun\, yr$^{-1}$}}

\newcommand{\rom}[1]{\uppercase\expandafter{\romannumeral#1}}

\newcolumntype{Y}{>{\centering\arraybackslash}X}

\usepackage[nameinlink,capitalise,noabbrev]{cleveref}

\crefname{section}{Section}{Sections}
\Crefname{section}{Section}{Sections}

\shorttitle{Infalling Streamers}
\shortauthors{Yun et al.}

\begin{document}

\title{Dynamical Impacts of Accretion Streamers on Protoplanetary Disks}

\author[0000-0003-4353-294X, gname='Han-Gyeol', sname='Yun']{Han-Gyeol Yun}
\affiliation{Department of Physics and Astronomy, Seoul National University, 1 Gwanak-ro, Gwanak-gu, Seoul 08826, Korea}
\email[show]{hangyeol@snu.ac.kr}

\author[0000-0003-3119-2087, gname='Jeong-Eun', sname='Lee']{Jeong-Eun Lee}
\affiliation{Department of Physics and Astronomy, Seoul National University, 1 Gwanak-ro, Gwanak-gu, Seoul 08826, Korea}
\affiliation{SNU Astronomy Research Center, Seoul National University, 1 Gwanak-ro, Gwanak-gu, Seoul 08826, Republic of Korea}
\email[show]{lee.jeongeun@snu.ac.kr}

\author[0000-0001-7258-770X, gname='Jaehan', sname='Bae']{Jaehan Bae}
\affiliation{Department of Astronomy, University of Florida, Gainesville, FL 32611, USA}
\email{jbae@ufl.edu}

\author[0000-0003-4625-229X, gname='Woong-Tae', sname='Kim']{Woong-Tae Kim}
\affiliation{Department of Physics and Astronomy, Seoul National University, 1 Gwanak-ro, Gwanak-gu, Seoul 08826, Korea}
\affiliation{SNU Astronomy Research Center, Seoul National University, 1 Gwanak-ro, Gwanak-gu, Seoul 08826, Republic of Korea}
\email{unitree@snu.ac.kr}

\begin{abstract}
Recent observations suggest that accretion streamers are common in protoplanetary disks, yet their dynamical impact on disk evolution remains poorly understood. Using three-dimensional hydrodynamic simulations with mass infall rates of $10^{-8}$--$10^{-6}\mdotyr$, we investigate how streamer accretion influences the structure and evolution of protoplanetary disks. We find that a prograde streamer can excite disk eccentricity globally to values as high as $\sim0.4$. The resulting eccentric disk develops prominent spiral arms and crescent-shaped overdensities whose spatial structures agree remarkably well with analytic eccentric-disk theory. It also undergoes significant warping, and exhibits enhanced and variable stellar accretion. In contrast, retrograde streamer accretion efficiently removes disk angular momentum, producing a compact disk and, in extreme cases, triggering disk breaking. Using synthetic ALMA molecular-line observations, we show that streamer-driven perturbations generate observable kinematic signatures, including Doppler flips in moment maps and wiggles in position--velocity diagrams. Remarkably, these signatures can persist for up to $100\kyr$ after infall has ceased, suggesting that some kinematic disturbances observed in disks without currently detected streamers may be relics of past infall events. Finally, we discuss the implications of streamer-driven disk evolution for planet formation and planet--disk interactions.
\end{abstract}



\section{Introduction}

In contrast to the classical picture of star formation, which envisioned protostellar systems as isolated objects undergoing gravitational collapse \citep{L1969, S1977, TS1984}, recent scattered-light and molecular-line observations have increasingly highlighted the importance of environmental interactions (see \citealt{PA2023}). These interactions often manifest as ``accretion streamers'', filamentary structures of infalling gas that connect protoplanetary disks (PPDs) to their surrounding natal material.

Accretion streamers have been observed in a wide variety of systems across different evolutionary stages. Among Class 0 protostars, Per-emb-2 \citep{PS2020} was found to host a large-scale streamer extending over $\gtrsim 10^4\au$, detected through molecular-line observations. Among Class I objects, ALMA observations revealed a streamer in HL Tau \citep{YG2019}, a system in which ongoing planet formation has been inferred from the presence of rings and gaps in the disk \citep{ALMA2015, DP2015}. Streamer-like infall has also been discussed in the eruptive young star V883 Ori, where asymmetric molecular structures connecting the outer disk/envelope to the disk were interpreted as a possible remnant or ongoing infall feature that may have perturbed the disk and contributed to triggering the FUor-like accretion outburst \citep{Lee2024}.
In addition, recent observations of L1489 IRS \citep{TL2024} and [BHB2007]-1 \citep{AC2020, GH2026} have suggested a possible connection between streamer infall and disk warping. Even during the more evolved Class II stage, when infall from the natal envelope is generally expected to have ceased, evidence for late-stage infall from accretion streamers has been reported in systems such as SU Aur \citep{AV2019}, DG Tau \citep{GP2022}, and AB Aur \citep{SD2025}. Furthermore, accretion streamers are not restricted to single-star systems: \citet{LMK2023} reported the discovery of three spiral arms acting as accretion streamers that funnel material onto the triple protostellar system IRAS 04239+2436. Recently, \citet{GB2026} found the streamer around GW Ori's misaligned disk has an angular momentum vector closely aligned with that of the outer disk, suggesting the streamer could be the cause of the  misalignment in the system.

These observational results suggest that accretion-streamer infall may be ubiquitous throughout star formation, underscoring the need to understand its impact on protostellar systems. Previous numerical studies have shown that interactions between disks and their large-scale environments can lead to the formation of a secondary disk \citep{DK2019}, the excitation of spiral structures \citep{KG2020}, and the perturbation of disk gas orbits, resulting in misaligned and eccentric outer disks \citep{KD2021}. However, these studies were based on the cloudlet-capture scenario, in which an initially spherical gas cloud is tidally stretched as it accretes onto the disk. More recent studies by \citet{CP2025, CP2025b} and \citet{HK2026} demonstrated that accretion streamers can likewise excite spiral structures and modify disk gas orbits.  Nevertheless, a more systematic investigation is needed to determine how these effects depend on the streamer's physical properties and to properly interpret future observations. 

In this paper, we investigate the dynamical impact of accretion streamers on PPDs using three-dimensional hydrodynamic simulations. To model the infalling material, we adopt a modified version of the Ulrich--Cassen--Moosman solution \citep{U1976, CM1981}, which lets us systematically control the streamer's geometry and mass infall rate. We first examine a set of fiducial models with mass infall rates ranging from $10^{-8}$ to $10^{-6}\mdotyr$ and characterize the resulting changes in disk structure, angular momentum distribution, and stellar accretion. We then perform synthetic molecular-line observations to identify observable signatures of streamer-driven perturbations and assess their detectability with ALMA.

To explore how disk evolution depends on streamer properties, we also consider models with different infall zones, allowing us to examine how the streamer's vertical position affects disk eccentricity and warping. We further investigate systems containing multiple streamers and compare their impact with that of a single streamer. Finally, motivated by recent observations and numerical studies suggesting retrograde streamers, we explore how counter-rotating infall alters disk evolution relative to the prograde case. Based on these results, we discuss the implications of streamer accretion for planet formation and planet--disk interactions.

This paper is organized as follows. In \autoref{sec:method}, we describe the numerical models and parameters adopted to investigate the impact of accretion streamers on PPDs. In \autoref{sec:hydro}, we present the hydrodynamic evolution of our fiducial streamer models, focusing on disk eccentricity (\S\ref{subsec:ecc}), substructures (\S\ref{subsec:substr}), angular momentum (\S\ref{subsec:angmom}), and stellar accretion (\S\ref{subsec:accrate}). In \autoref{sec:synobs}, we examine how these streamer-induced perturbations manifest in synthetic ALMA molecular-line observations. Specifically, we investigate their observational signatures using velocity channel maps (\S\ref{subsec:channel}), moment maps (\S\ref{subsec:moment}), and position--velocity diagrams (\S\ref{subsec:pv}). In \autoref{sec:discuss}, we explore the effects of the streamer's vertical position (\S\ref{subsec:vpos}), multiplicity (\S\ref{subsec:multi}), and retrograde motion (\S\ref{subsec:retro}), and discuss the broader implications (\S\ref{subsec:implication}) and limitations (\S\ref{subsec:caveat}) of our study. Finally, we summarize our main results in \autoref{sec:summary}.

\section{Method} \label{sec:method}
\subsection{Basic Equations}
In this paper, we model a PPD as a three-dimensional viscous fluid, neglecting self-gravity and magnetic fields. We use FARGO3D \citep{BM2016} to solve the hydrodynamic equations of mass, momentum, and energy conservation in spherical polar coordinates $(r, \theta, \phi)$:
\begin{align}
    \frac{\partial\rho}{\partial t} + \nabla \cdot (\rho \mathbf{v} ) &= 0\,, \label{eq:cont} \\
    \frac{\partial\mathbf{v}}{\partial t} + (\mathbf{v}\cdot\mathbf{\nabla}) \mathbf{v} &= -\frac{\nabla P}{\rho}-\nabla\Phi-\frac{\nabla\cdot {\Pi}}{\rho}\,, \label{eq:moment} \\
    \frac{\partial e}{\partial t} + \nabla \cdot (e \mathbf{v} )  &= - P \mathbf{\nabla}\cdot\mathbf{v} + Q_{\text{visc}} -Q_{\text{cool}}\,. \label{eq:energy}
\end{align}
Here, $\rho$, $\mathbf v=(v_r,v_\theta,v_\phi)$, $P$, and $e$ denote the gas density, velocity, pressure, and internal energy density, respectively. The pressure is related to the internal energy density through $P=(\Gamma-1)e$, where the adiabatic index is fixed to $\Gamma=1.4$.

In \cref{eq:moment}, $\Phi(r)=-GM_*/r$ is the gravitational potential of the central star with mass $M_*$. The quantity $\Pi$ denotes the viscous stress tensor, defined as
\begin{equation}\Pi\equiv\rho\nu\left[\nabla\mathbf{v}+(\nabla\mathbf{v})^T-\frac{2}{3}\left(\mathbf{\nabla}\cdot v\right)\mathbb{I}\right]\,, \label{eq:visctensor}
\end{equation}
where $\nu$ is the kinematic viscosity and $\mathbb{I}$ is the identity matrix. We parameterize the viscosity using the standard $\alpha$ prescription, $\alpha\equiv \nu \Omega_K/ c_s^2$ \citep{SS1973}, where $c_s$ and $\Omega_K=\sqrt{GM_*/r^3}$ are the isothermal sound speed and the Keplerian angular frequency, respectively. Throughout this work, we adopt $\alpha=10^{-4}$, which is sufficiently large to suppress the vertical shear instability \citep{NG2013}

In \cref{eq:energy}, we include three principal thermal processes: compressional heating ($-P\mathbf{\nabla}\cdot{\mathbf{v}}$), viscous dissipation ($Q_{\text{visc}}$) and radiative cooling ($Q_{\text{cool}}$). The viscous heating and the radiative cooling rates are given by
\begin{align}
    Q_{\text{visc}} &= \frac{1}{2\nu\rho}\Pi_{ij}\Pi^{ij}+\frac{2\nu\rho}{9}\left(\mathbf{\nabla}\cdot\mathbf{v}\right)^2\,, \label{eq:vischeat} \\
    Q_{\text{cool}} &= \rho c_V \frac{T-T_{\text{eq}}}{t_{\text{cool}}}\,, \label{eq:radcool}
\end{align}
where $c_V$ is the specific heat capacity at constant volume, $T\equiv \mu c_s^2/\mathcal{R}$ is the gas temperature assuming an ideal equation of state, $\mathcal{R}$ is the universal gas constant, and $\mu\simeq 2.4\,{\rm g\,mol^{-1}}$ is the mean molecular weight. \Cref{eq:radcool} describes Newtonian cooling, driving the gas temperature toward the equilibrium temperature $T_{\rm eq}$ on a characteristic cooling timescale $t_{\rm cool}$.

We follow \citet{LR2016} to estimate the cooling timescale $t_{\text{cool}}$. First, we determine the opacity $\kappa$ in each grid cell from the local density $\rho$ and temperature $T$ using the opacity prescription of \citet{ZH2009}. The optical depth $\tau$ at a cylindrical height $z=r\cos\theta$ is then computed from the optical depths toward the upper and lower disk surfaces as 
\begin{equation}
    \tau = \left(\frac{1}{\tau_{\text{upper}}} + \frac{1}{\tau_{\text{lower}}}\right)^{-1}\,,
\end{equation}
where
\begin{align}
    \tau_{\text{upper}} &= \int_{z}^{z_{\text{upper}}} \rho(z')\kappa(z')dz'\,, \\
    \tau_{\text{lower}} &= \int_{z_{\text{lower}}}^{z} \rho(z')\kappa(z')dz'\,.
\end{align}
Here, $z_{\mathrm{upper}}$ and $z_{\mathrm{lower}}$ denote the upper and lower disk surfaces, respectively. By interpolating between the optically thick and optically thin limits \citep{H1990, DH2003}, we define the effective optical depth as
\begin{equation}
    \tau_{\text{eff}} = \frac{3}{8}\tau + \frac{\sqrt{3}}{4} + \frac{1}{4\tau}\,.
\end{equation}
The cooling timescale $t_{\text{cool}}$ is estimated from the radiative diffusion timescale
\begin{equation}
    t_{\text{cool}} \equiv \frac{\int edV}{\int \mathbf{F} \cdot d\mathbf{A}}\,, \label{eq:coolint}
\end{equation}
where $\mathbf{F} = \hat{\mathbf{n}} \sigma T^4 / \tau_{\text{eff}}$, with $\sigma$ being the Stefan--Boltzmann constant and $\hat{\mathbf{n}}$ the unit vector normal to the surface element $d\mathbf{A}$. Evaluating \cref{eq:coolint} over a spherical volume with radius equal to the local pressure scale height, $H\equiv c_s/\Omega_K$, yields
\begin{equation}
    t_{\text{cool}} = \frac{\rho c_VH\tau_{\text{eff}}}{3\sigma T^3}\label{eq:tcool}
\end{equation}
for the cooling timescale. In the disk mid-plane, the dimensionless cooling time is $t_{\mathrm{cool}}\Omega_K$ $\sim4$ at
$R=300\,\mathrm{au}$ and $\sim0.6$ at $R=100\,\mathrm{au}$, indicating that
radiative cooling is relatively efficient throughout the disk.

\subsection{Disk Setup}

As the initial disk model, we adopt the temperature profile introduced by \citet{DD203}:
\begin{align}
    &T(R,z) = \notag \\
    &\begin{cases}
        T_{\text{mid}}(R) + \left[T_{\text{atm}}(R)-T_{\text{mid}}(R)\right]\sin^2\left(\frac{\pi z}{2z_q}\right)\,, & \text{if } z<z_q\,, \\
        T_{\text{atm}}(R)\,, & \text{if } z\geq z_q\,,
    \end{cases}
\end{align}
where $R\equiv r\sin\theta$ is the cylindrical radius. The midplane and atmospheric temperatures are given by 
$T_{\text{mid}}(R)=T_{\text{mid},R_c}(R/R_c)^{-q_T}$ and $T_{\text{atm}}=T_{\text{atm},R_c}(R/R_c)^{-q_T}$, respectively. Here, $R_c$ is a characteristic radius and $q_T$ is the radial temperature exponent. The transition height is set to  $z_q=4H_{\mathrm{mid}}$, where 
\begin{equation}
    H_{\mathrm{mid}} = \frac{c_s}{\Omega_K} =\sqrt{\mathcal{R}\frac{T_{\text{mid}}}{\mu}}\Omega_K^{-1}
\end{equation}
is the pressure scale height at the disk midplane. 
The resulting temperature structure features a cool midplane embedded within a warmer atmosphere.

The initial density distribution is obtained by assuming vertical hydrostatic equilibrium,
\begin{equation}
    \rho(R,z) = \rho_{\text{mid}}(R)\frac{c_s^2(R,0)}{c_s^2(R,z)}\exp\left[-\int_0^z\frac{1}{c_s^2(R,z')}\frac{\partial\Phi}{\partial z'}dz'\right]\,.
\end{equation}
The midplane density profile $\rho_{\text{mid}}(R)$ is  determined from the prescirbed surface density profile
\begin{equation}
    \Sigma(R) = \Sigma_{R_c}\left(\frac{R}{R_c}\right)^{-q_{\Sigma}}\,,
\end{equation}
where $\Sigma_{R_c}$ is the surface density at the characteristic radius $R_c$ and $q_{\Sigma}$ is the radial power-law index of the surface density.

Finally, assuming $v_R=v_z=0$, the azimuthal velocity is determined from radial force balance:
\begin{equation}
    v_{\phi}(R,Z)=\left[\frac{GM_*R^2}{(R^2+z^2)^{3/2}}+\frac{R}{\rho}\frac{\partial P}{\partial R}\right]^{1/2}.
\end{equation}
The resulting density, temperature, and velocity distributions serve as the initial conditions for all simulations.

\begin{table*}
    \centering
    \begin{tabular}{c c c}
        \hline\hline
        Notation & Description &  Value \\
        \hline
        $M_*$ & Stellar mass & 2.4 $\msun$ \\
        $R_c$ & Characteristic radius & 98 au \\
        $q_T$   & Temperature power-law index & 0.1 \\
        $T_{\text{mid},R_c}$ & Mid-plane temperature at $R_c$ & 42 K \\
        $T_{\text{atm},R_c}$ & Atmosphere temperature at $R_c$ & 70 K \\
        $q_{\Sigma}$ & Surface density power-law index & 2.15 \\
        $\Sigma_{R_c}$ & Surface density at $R_c$ & 0.5 g cm$^{-2}$ \\
        \hline
    \end{tabular}
    \caption{Initial disk parameters} \label{tbl:diskparam}
\end{table*}

\begin{figure*}[htb!]
    \epsscale{1.15}
    \plotone{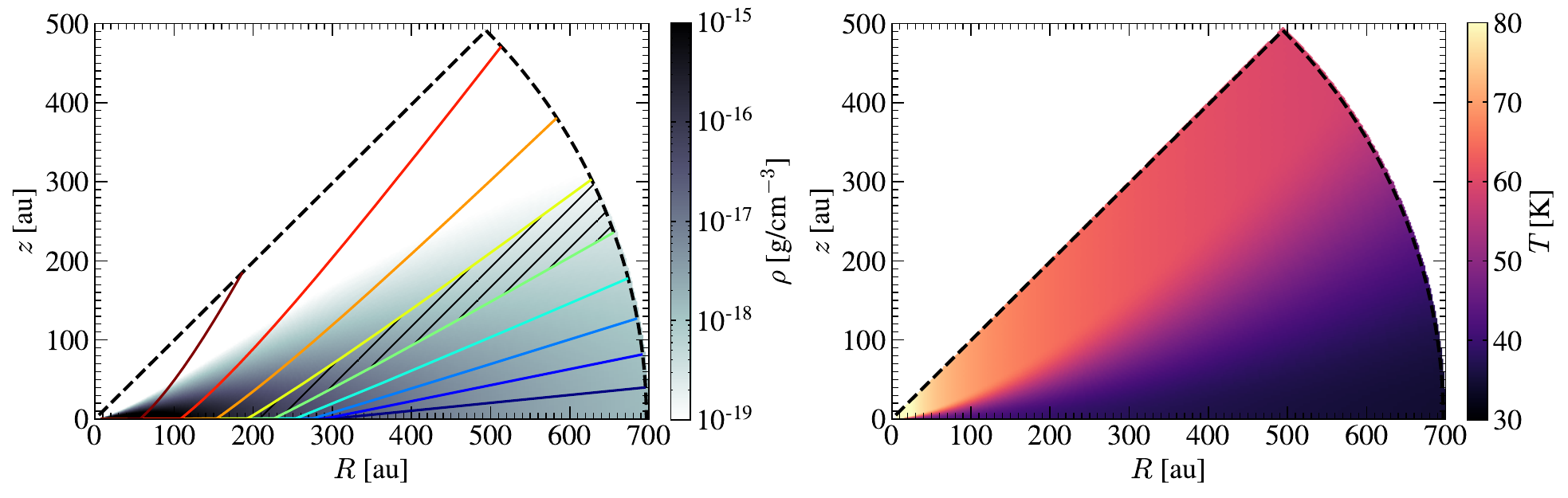}
    \caption{Two-dimensional distributions of (\textit{left}) the density $\rho$  and (\textit{right}) temperature $T$  in the $R-z$ plane. 
    In the left panel, streamlines of infalling material are overlaid for $\cos\theta_0$ ranging from 0.1 to 0.9 in steps of 0.1, with values increasing from bottom to top, assuming a centrifugal radius of $R_{\mathrm{cr}}=300\,\mathrm{au}$. The hatched region indicates the infall zone of the fiducial models in this study.}
    \label{fig:diskmodel}
\end{figure*}

The parameters adopted for our disk model are summarized in \cref{tbl:diskparam}. We adopt values based on the astrochemical model of AB Aur presented by \citet{RF2020}. AB Aur has recently been identified as a system undergoing late-stage infall \citep{SD2025}, making it a suitable target for the present study. The resulting density and temperature distributions are shown in \cref{fig:diskmodel}.

\subsection{Infall Model}
In this study, we model the infalling material using the classical collapse solution developed by \citet{U1976} and \citet{CM1981}, hereafter the UCM model, with several modifications. The UCM model describes the density and velocity structure of a rotating, collapsing envelope in a quasi-steady state. It assumes that gas parcels move along ballistic trajectories under the gravitational influence of the central protostar while conserving both angular momentum and total mechanical energy. Under these assumptions, the velocity field of the infalling gas is given by 
\begin{align}
    v_r(r,\theta) &= -r\Omega_K\left(1+\frac{\cos\theta}{\cos\theta_0}\right)^{1/2}\,, \label{eq:ucmvr} \\
    v_{\theta}(r,\theta) &= r\Omega_K\left(\frac{\cos\theta_0-\cos\theta}{\sin\theta}\right)\left(1+\frac{\cos\theta}{\cos\theta_0}\right)^{1/2}\,, \label{eq:ucmvtheta} \\
    v_{\phi}(r, \theta) &= r\Omega_K\left(\frac{\sin\theta_0}{\sin\theta}\right)\left(1-\frac{\cos\theta}{\cos\theta_0}\right)^{1/2} \label{eq:ucmvaz}\,,
\end{align}
where $\theta_0(r, \theta)$ denotes the initial polar angle of a gas parcel currently located at $(r,\theta)$. Surfaces of constant $\theta_0$ therefore define the streamlines of the infalling gas.

To determine the distribution of $\theta_0$, we specify the centrifugal radius $R_{\text{cr}}$, defined as the radius in the equatorial plane where the centrifugal and gravitational forces are in balance. We adopt $R_{\text{cr}}=300\au$, corresponding to the location of the merging zone identified in AB Aur by \citet{SD2025}. 

The value of $\theta_0$ is obtained from the streamline equation. Defining $\xi(r)\equiv R_{\text{cr}}/r$, one obtains 
\begin{equation}
    \xi(r)\cos^3\theta_0 + \left[1-\xi(r)\right]\cos\theta_0-\cos\theta=0\,,
\end{equation}
which is a depressed cubic equation for $\cos\theta_0$ and can therefore be solved analytically. The resulting streamlines, corresponding to constant values of  $\cos\theta_0$ for $R_{\mathrm{cr}}=300$ au, are illustrated in the left panel of \cref{fig:diskmodel}.

In the UCM model, the density distribution of the infalling material is given by
\begin{equation}
    \rho_{\text{UCM}}(r,\theta) = -\frac{\dot{M}_{\mathrm{in}}}{4\pi r^2v_r}\frac{1}{1+\xi(r)(2-3\sin^2\theta_0)}\,,
\end{equation}
where $\dot{M}_{\mathrm{in}}$ is the mass infall rate. This expression assumes isotropic infall over the entire solid angle. In our streamer model, however, the infall is confined to a limited range of polar and azimuthal angles. If the infall zone is restricted to $\cos\theta_0 \in\left[\cos\theta_{0,\text{min}},\,\cos\theta_{0,\text{max}}\right]$, with an azimuthal extent of $\Delta\phi$, its solid angle is 
\begin{equation}
    {\Delta \Omega_\text{infall}} = \left|\cos\theta_{0,\text{max}}-\cos\theta_{0,\text{min}}\right|\Delta\phi\,.
\end{equation}
To preserve the prescribed mass infall rate within this restricted region, we scale the density according to
\begin{equation}
    \rho_{\text{in}} = \rho_{\text{UCM}}\left(\frac{{\Delta \Omega_\text{infall}}}{4\pi}\right)^{-1}\,, \label{eq:ucmdens}
\end{equation}
which reduces to the standard UCM density distribution when the infall zone covers the entire sphere. 

In our fiducial models, the infall zone is restricted to $\cos\theta_0\in[0.5,0.6]$, such that infall occurs only above the disk midplane. We further adopt an azimuthal extent of $\Delta\phi=0.4$, with the streamer centered at $\phi=0$.
Unless otherwise noted, the results presented in \autoref{sec:hydro} and \autoref{sec:synobs} are based on these fiducial models. While our choice of $\Delta\phi=0.4$ is somewhat arbitrary, we find that varying the azimuthal extent from $\Delta\phi=0.2$ to $0.6$ produces no notable differences in the results. We also perform supplementary simulations with different ranges of $\cos\theta_0$ to investigate the effect of the streamer's vertical location (\autoref{subsec:vpos}).

To explore the impact of streamer accretion on disk evolution, we consider three mass infall rates $\dot{M}_{\mathrm{in}}=10^{-8}, 10^{-7}$, and $10^{-6}\mdotyr$. Each simulation begins with a 2 kyr relaxation phase without infall ($t=-2$ to 0 kyr), allowing the disk to settle into a new quasi-equilibrium state established by the balance between viscous heating, compressional heating, and radiative cooling. At the disk characteristic radius $R_c$, the local orbital period is $2\pi/\Omega_K\sim 0.63\kyr$, while the thermal timescale is $\sim 0.6\Omega_K^{-1}\sim 0.06\kyr$. Thus, the $2\kyr$ relaxation phase spans more than three orbital periods and about 30 thermal timescales, providing sufficient time for the disk to settle into a quasi-equilibrium state.

We then apply infall for 8 kyr ($t=0$ to 8 kyr), followed by a 92 kyr evolution period without infall ($t=8$ to 100 kyr).
This setup enables us to assess both the immediate response of the disk to streamer accretion and the long-term evolution of the induced structures  after infall ceases.

\subsection{Numerical Setup}

In our simulations, the computational domain in spherical polar coordinates $(r, \theta, \phi)$ extends from $7$ to $700$ au in the radial direction, from $\pi/4$ to $3\pi/4$ in the meridional direction, and over the full azimuthal range $0\le\phi<2\pi$. The grid consists of 256 logarithmically spaced cells in $r$, 384 uniformly spaced cells in $\phi$ chosen to satisfy $\Delta r\sim r\Delta\phi$, and 128 uniformly spaced cells in $\theta$. At the disk mid-plane, the pressure scale height $H_{\textrm{mid}}$ at $R_c$ is resolved by approximately seven grid cells.

At the outer radial boundary within the infall region, the velocity and density fields are fixed to the UCM solution given by  \cref{eq:ucmvr,eq:ucmvtheta,eq:ucmvaz,eq:ucmdens}. Outside the infall region, zero-gradient boundary conditions are applied to all variables at the radial boundaries, except for $v_r$, for which an outflow boundary condition is adopted. In this case, we set ghost-zone values equal to those of the last active cell only when the flow is directed out of the computational domain. To reduce spurious wave reflections, we additionally implement a wave-damping zone \citep{VB2006} near the inner radial boundary over the range $r=7$--$9.2$ au, where $\rho$ and $v_{\phi}$ are relaxed toward their initial profiles.

At the meridional boundaries, zero-gradient boundary conditions are imposed on $v_r, v_{\phi}$, and $T$, while an outflow boundary condition is applied to $v_{\theta}$. For the density, the ghost-zone values are determined by enforcing hydrostatic equilibrium in the meridional direction,
\begin{equation}
    \frac{1}{\rho}\frac{\partial}{\partial\theta}(\rho c_s^2)=\frac{v_{\phi}^2}{\tan\theta}\,.
\end{equation}
This treatment helps maintain vertical hydrostatic equilibrium near the meridional boundaries. Finally, because the computational domain spans the full azimuthal range, periodic boundary conditions are applied in the $\phi$ direction.

\section{Simulation Results} \label{sec:hydro}

\begin{figure*}[htb!]
    \epsscale{1.0}
    \plotone{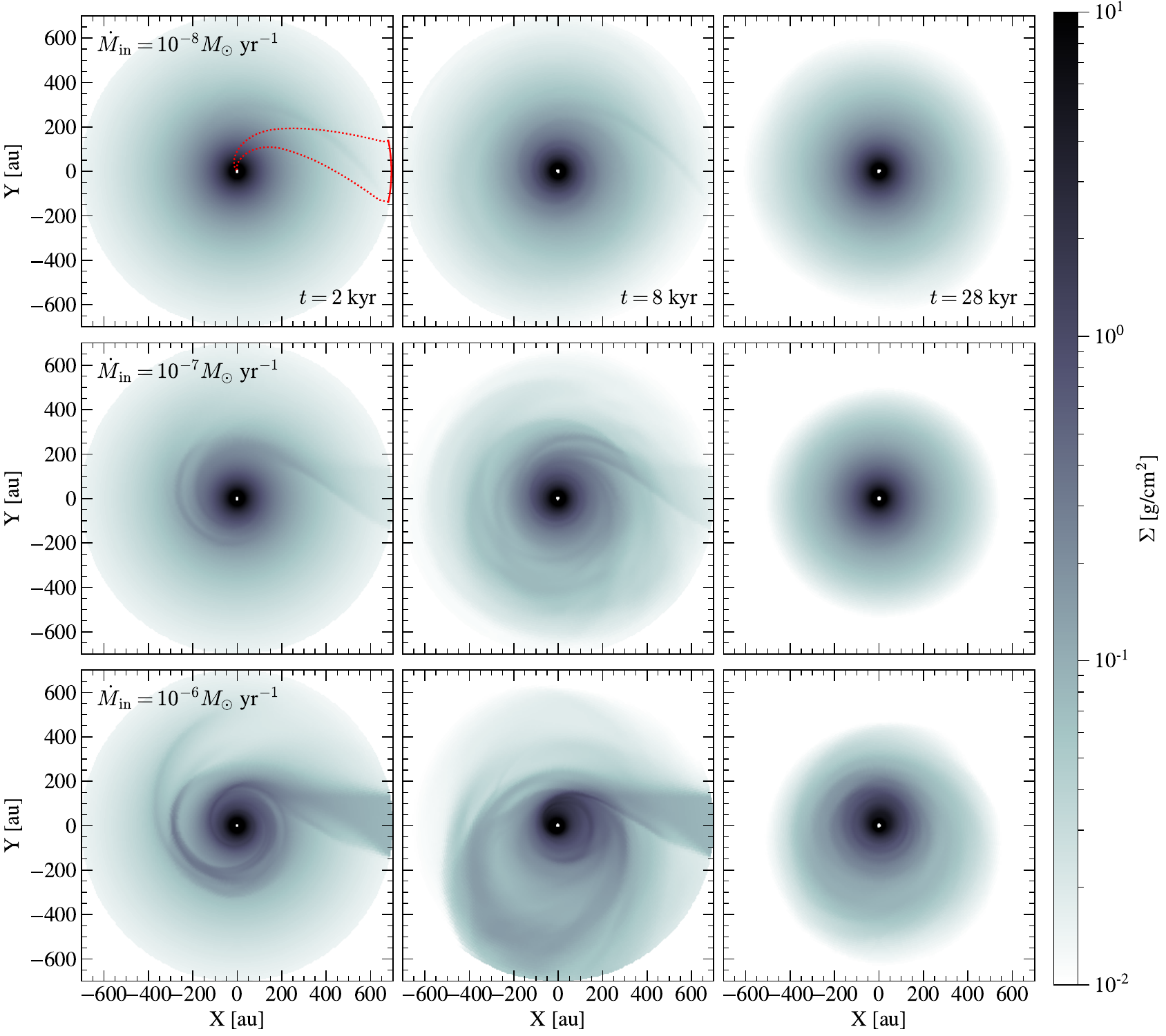}
    \caption{Snapshots of the disk surface density at $t=2\kyr$ (\textit{left}), $8\kyr$ (\textit{middle}) and $28\kyr$ (\textit{right}), for infall rates of $\dot{M}_{\mathrm{in}}=10^{-8}\mdotyr$ (\textit{top}), $10^{-7}\mdotyr$ (\textit{middle}) and $10^{-6}\mdotyr$ (\textit{bottom}). The short red arc and red dotted lines in the upper-left panel mark the infall zone and the infall trajectories of the fiducial model, respectively.}
    \label{fig:spiralsnapshot}
\end{figure*}

In this section, we present the PPD's hydrodynamic response to ongoing infall from an accretion streamer. Shortly after the infalling material from the accretion streamer collides with the outer disk, the velocity perturbations it induces generate various disk substructures that extend throughout the entire disk. At the same time, the asymmetric, parabolic shape of the streamer excites global disk
eccentricity, rendering the overall disk structure highly non-axisymmetric. \cref{fig:spiralsnapshot} provides a comprehensive overview of this evolution, presenting the disk surface density at multiple snapshots
($t=2$, 8, and 28\,kyr) for varying mass infall rates
($\dot{M}_{\mathrm{in}} = 10^{-8}$ --  $10^{-6}\mdotyr$). These panels show
that the accretion streamer has significant morphological and kinematic effects on the disk. In what follows, we organize our discussion of
these effects around four primary physical outcomes: (1) disk eccentricity, (2) disk substructures, (3) disk angular momentum, and
(4) the accretion rate.

\subsection{Disk Eccentricity} \label{subsec:ecc}

\begin{figure*}[htb!]
    \epsscale{1.0}
    \plotone{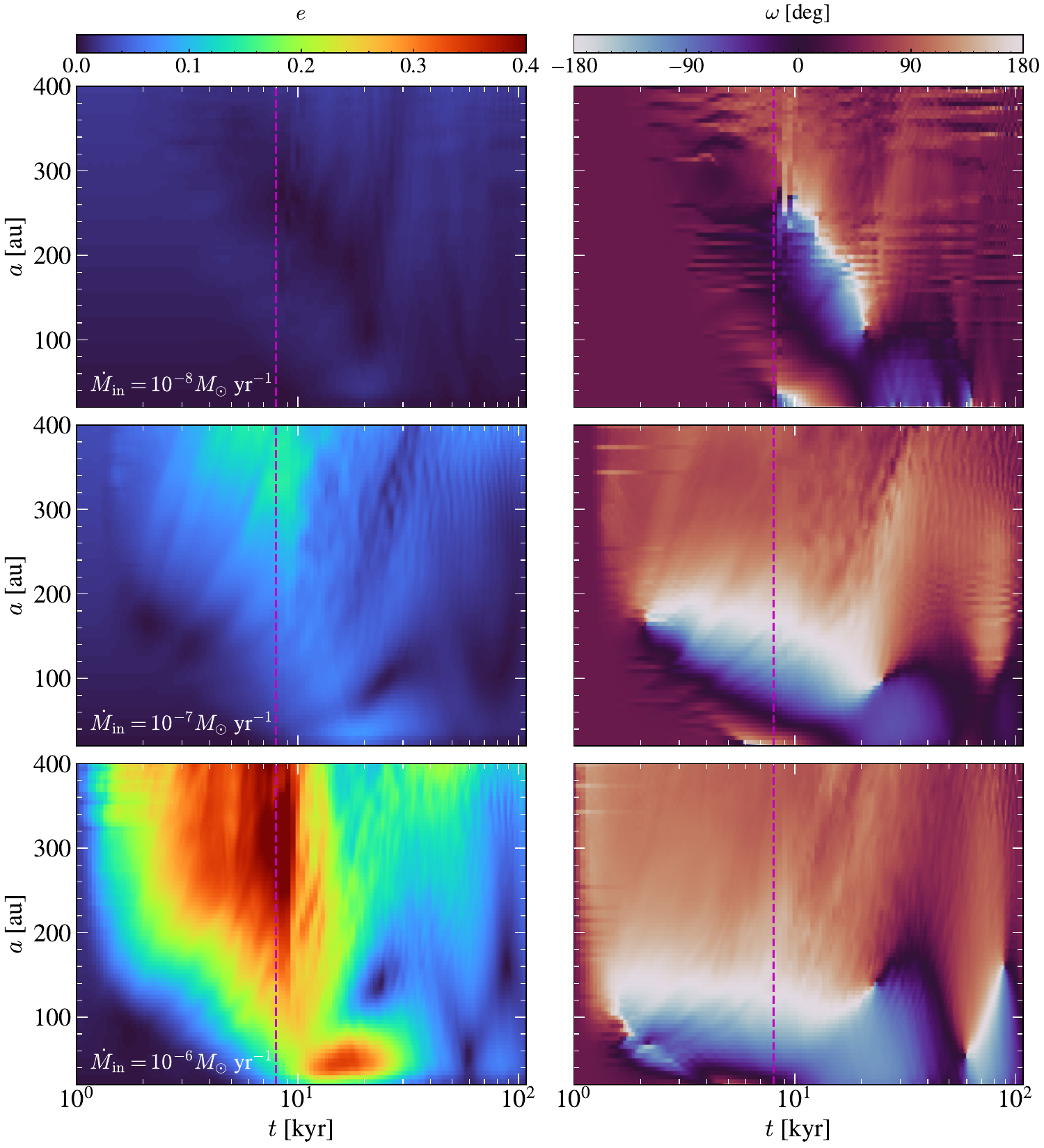}
    \caption{Temporal evolution of the disk eccentricity $e$ (\textit{left}) and the argument of periapsis $\omega$ (\textit{right}), each averaged azimuthally at fixed semi-major axis $a$, for mass infall rates of $\dot{M}_{\mathrm{in}}=10^{-8}\mdotyr$ (\textit{top}), $10^{-7}\mdotyr$ (\textit{middle}) and $10^{-6}\mdotyr$ (\textit{bottom}). The period during which infall is halted is indicated by the vertical dashed lines. In all cases, the streamer excites the eccentricity in the outer disk, which then propagates inward. After infall ceases, the eccentricity damps but remains elevated in the inner disk, while the radial gradient of $\omega$ drives the ``twisting" of the disk.}
    \label{fig:orbitalelements}
\end{figure*}

The primary impact of the streamer on the disk is the excitation of the disk eccentricity by the parabolic streamer \citep{CP2025}. To investigate the effect of the accretion streamer on the orbits of the disk gas, we measure the orbital elements of the disk following the method described in \citet{TO2017}, which we briefly summarize here. First, we calculate the various orbital elements of each cell from its position and velocity vectors. We then group the cells into radial bins of the semi-major axis $a$, with a bin width of $\delta a=10\au$. Finally, we compute the density-weighted average of the eccentricity $e$ and the argument of periapsis $\omega$ in each bin, which together represent the azimuthally averaged and vertically integrated orbital shape and orientation of an elliptical orbit with semi-major axis $a$.

\cref{fig:orbitalelements} displays the evolution of the disk eccentricity ($e$) and the argument of periapsis ($\omega$) for mass infall rates of $\dot{M}_{\mathrm{in}}=10^{-8}, 10^{-7}$, and $10^{-6}\mdotyr$. We observe that the disk eccentricity is initially excited in the outer disk and subsequently propagates inward. Although the lowest infall rate ($\dot{M}_{\mathrm{in}}=10^{-8}\mdotyr$) is insufficient to drive significant disk eccentricity, the streamer with an infall rate $\dot{M}_{\mathrm{in}}=10^{-7}\mdotyr$ excites the disk eccentricity to $\sim 0.04$ in the inner disk ($R\sim100\au$) and $0.12$ in the outer disk ($R\sim 300\au$) by the end of the infall period ($t=8$ kyr). At the higher infall rate ($\dot{M}_{\mathrm{in}}=10^{-6}\mdotyr$), the disk eccentricity reaches $\sim0.24$ in the inner disk and $\sim0.42$ in the outer disk by the end of the infall. Concurrently, a radial gradient in $\omega$ develops during infall, causing the disk to become ``twisted" \citep{OB2014}. 

Once infall ceases, the outer-disk eccentricity begins to damp. The eccentricity in the inner disk, however, continues to grow until $t\sim 20\kyr$ as the eccentricity from the outer disk propagates inward, reaching $\sim0.07$ and $\sim0.3$ at $a\sim 40\au$ for $\dot{M}_{\mathrm{in}}=10^{-7}\mdotyr$ and $\dot{M}_{\mathrm{in}}=10^{-6}\mdotyr$, respectively. The eccentricity subsequently damps across the entire disk, although we simultaneously observe an oscillation in the disk eccentricity, with the eccentricity variations in the inner and outer disk found to be anti-correlated. 

\begin{figure*}[htb!]
    \epsscale{1.0}
    \plotone{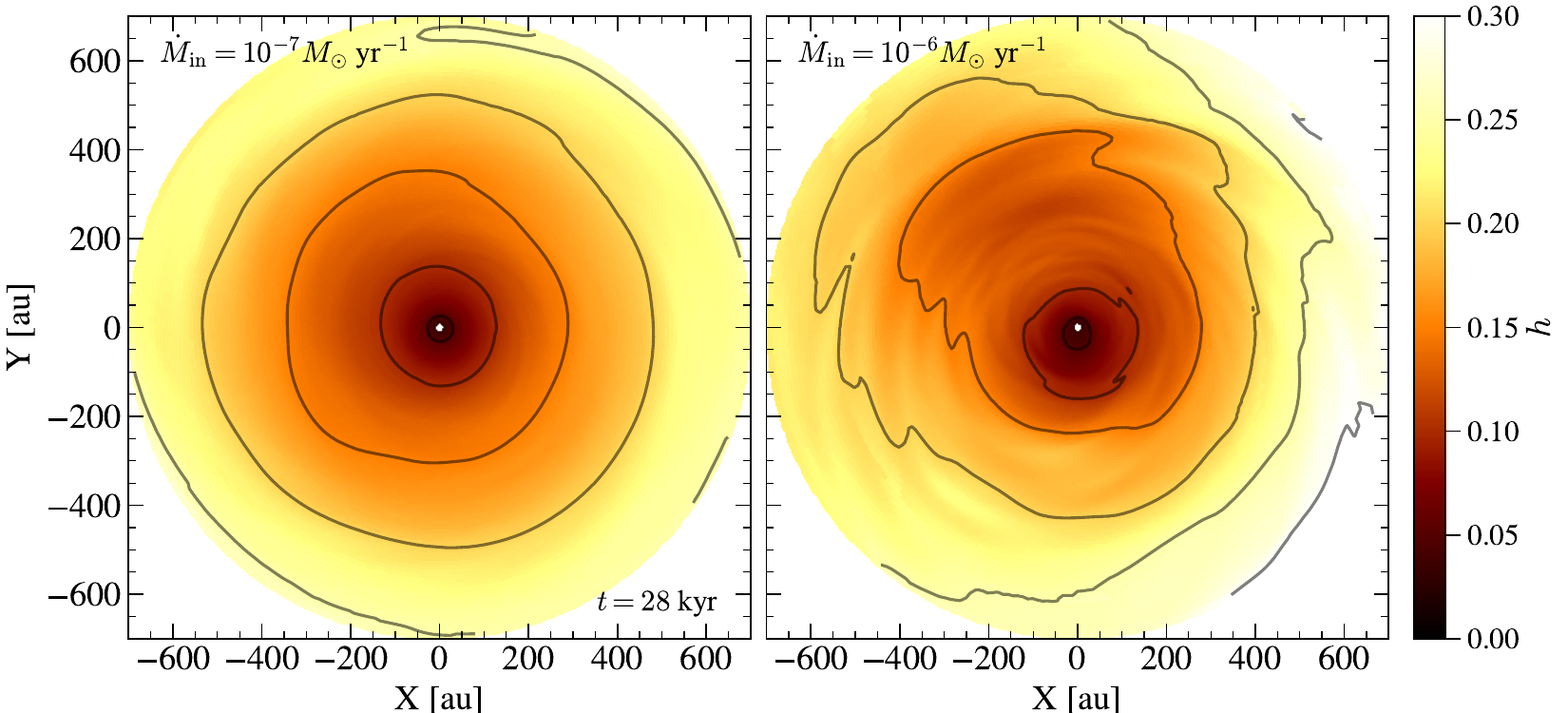}
    \caption{Disk aspect ratio $h\equiv H/R$ measured $20\kyr$ after the infall has stopped ($t=28\kyr$), for $\dot{M}_{\mathrm{in}}=10^{-7}\mdotyr$ (\textit{left}) and $\dot{M}_{\mathrm{in}}=10^{-6}\mdotyr$ (\textit{right}). The contour lines correspond to $h=0.05$--$0.30$, in increments of 0.05. As the streamer excites greater eccentricity, the disk aspect ratio becomes increasingly non-axisymmetric, persisting even $\sim 20\kyr$ after the infall has ended.}
    \label{fig:aspectratio}
\end{figure*}

This behavior is consistent with the eccentricity evolution described in previous theoretical \citep{OB2014, OL2019} and numerical studies \citep{CL2024, Deng2026}. The eccentricity initially excited in the outer disk propagates inward as a pressure-driven dispersive wave. During this redistribution, the disk develops a negative eccentricity gradient, which reduces the likelihood of orbital intersections \citep{CL2024, Deng2026}. Differential apsidal precession causes the apsidal twist $\mathrm{d}\omega/\mathrm{d}a$ to change sign periodically, thereby reversing the direction of angular momentum deficit flux \citep{OL2019}. This periodic reversal drives the eccentricity oscillations observed in our simulations \citep{Deng2026}. 
A quantitative analysis of the eccentricity decay at the post-infall phase is presented in Appendix \ref{app:amd}.

A major consequence of an eccentric disk is the development of a non-axisymmetric vertical structure. This arises from vertical oscillations driven primarily by gas infall and is further modulated by variations in the vertical gravity along eccentric orbits \citep{OB2014, RL2024}. To quantify this effect, we measure the disk scale height $H$ using the second moment of the density distribution:
\begin{equation}
    H = \left( \frac{\int z^2\rho dz}{\int \rho dz} \right)^{1/2}\,.
\end{equation}
With this definition, the initial disk has a scale height of $H\simeq 0.32 (R/R_0)^{1.45}\au$. \cref{fig:aspectratio} plots the disk aspect ratio, $h\equiv H/R$, measured at $t=28\kyr$, after the infall has ceased, for modes with $\dot{M}_{\mathrm{in}}=10^{-7}\mdotyr$ and $10^{-6}\mdotyr$. In both cases, the aspect ratio exhibits clear azimuthal variations associated with the eccentric disk geometry. At $R=300\au$, the fractional deviation from the azimuthal average reaches $\sim 1\%$ for the $\dot{M}_{\mathrm{in}}=10^{-7}\mdotyr$ model and $\sim 4\%$ for the $\dot{M}_{\mathrm{in}}=10^{-6}\mdotyr$ model. The corresponding peak-to-peak variations in scale height are $\sim 6.5\au$ and $\sim 22.5\au$, respectively.

To test whether the scale height variation is driven by the eccentricity, we compare our results with the analytic framework of \citet{OL2019}. They showed that the geometric compression and expansion associated with eccentric orbits can be characterized by the parameter $j$, defined as
\begin{equation}
    j = \frac{1-e(e+a\,de/da)}{\sqrt{1-e^2}} \left[1-q\cos(E-\chi)\right]\,, \label{eq:OL_j}
\end{equation}
where $E$ is the eccentric anomaly, while the amplitude $q$ and the phase angle $\chi$ are given by
\begin{align}
    q\cos\chi &= \frac{a\,de/da}{1-e(e+a\,de/da)}\,, \\
    q\sin\chi &= \frac{\sqrt{1-e^2}ae\,d\omega/da}{1-e(e+a\,de/da)}\,.
\end{align}
The scale height $H$ of the eccentric disk can then be expressed relative to that of a reference circular disk, $H_o$, through the dimensionless modulation factor $\lambda_H \equiv H/H_o$. The variation of $\lambda_H$ along an eccentric orbit is obtained by solving 
\begin{equation}
    (1-e\cos E)\frac{d^2\lambda_H}{dE^2}-e\sin E\frac{d\lambda_H}{dE} + \lambda_H = \frac{(1-e\cos E)^3}{j^{\Gamma-1}\lambda_H^{\Gamma}}\,. \label{eq:OL1}
\end{equation}

\begin{figure*}[htb!]
    \epsscale{0.9}
    \plotone{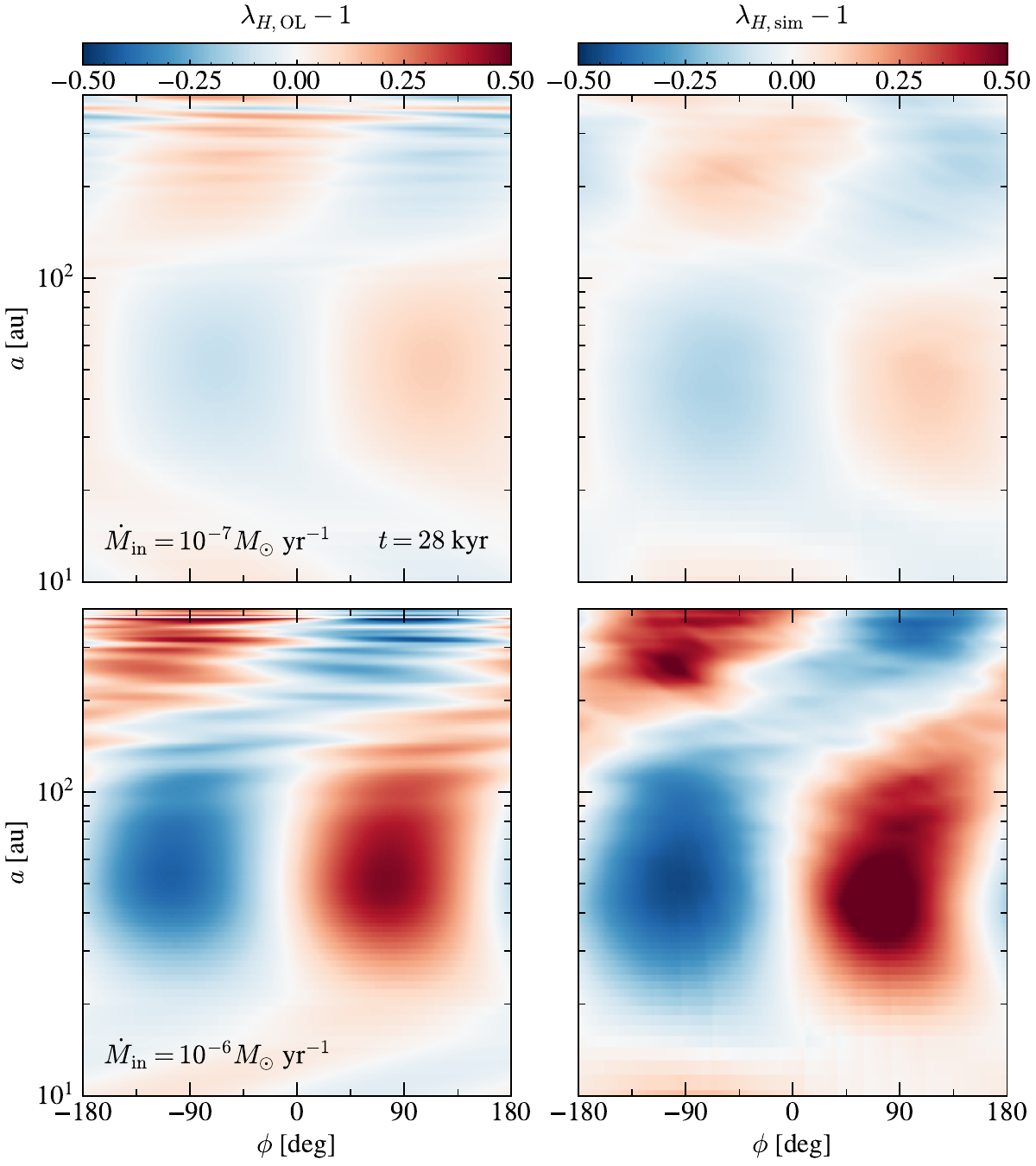}
    \caption{Scale height modulation factor $\lambda_H$ obtained by solving \cref{eq:OL1} ($\lambda_{H,\mathrm{OL}}$, \textit{left}) and from the simulations ($\lambda_{H,\mathrm{sim}}$,\textit{right}) for $\dot{M}_{\mathrm{in}}=10^{-7}\mdotyr$ (\textit{top}) and $10^{-6}\mdotyr$ (\textit{bottom}) at $t=28\kyr$. The simulated scale height variations agree well with those predicted by the analytic framework of \citet{OL2019}.}
    \label{fig:OL_scaleheight}
\end{figure*}

\cref{fig:OL_scaleheight} compares the scale height modulation factor predicted by \cref{eq:OL1}, $\lambda_{H,\mathrm{OL}}$, with that measured from the simulations, $\lambda_{H,\mathrm{sim}}$, for $\dot{M}_{\mathrm{in}}=10^{-7}\mdotyr$ and $10^{-6}\mdotyr$ at $t=28\kyr$. At a given semi-major axis $a$, the analytic model qualitatively reproduces both the phase and amplitude of the azimuthal scale height variation seen in the simulations. The remarkable agreement between the analytic prediction and simulation indicates that the non-axisymmetric vertical structure of the disk is primarily driven by the eccentricity excited by the accretion streamer.

\subsection{Disk Substructures} \label{subsec:substr}

\begin{figure*}[htb!]
    \epsscale{1.0}
    \plotone{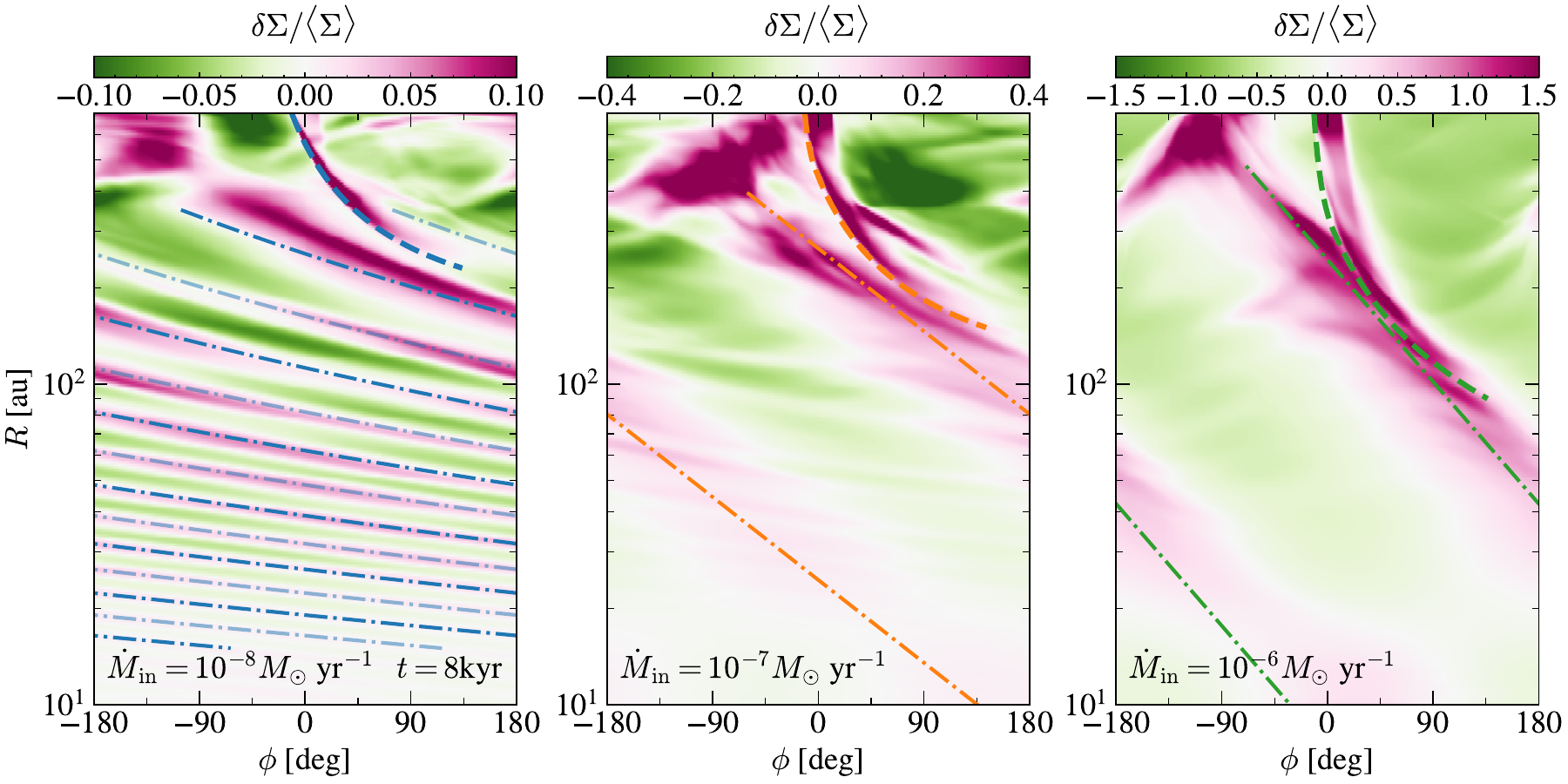}
    \caption{Surface density perturbation $\delta\Sigma/\langle\Sigma\rangle$, measured at the end of the infall phase ($t=8\kyr$) for $\dot{M}_{\mathrm{in}}=10^{-8}$ (\textit{left}), $10^{-7}$ (\textit{middle}), and $10^{-6}\mdotyr$ (\textit{right}). 
    Dashed curves at large radii trace the primary spiral arm, while dot-dashed lines at smaller radii indicate the locations of the secondary spiral arm.}
    \label{fig:primaryfit}
\end{figure*}

Another major consequence of the streamer-disk interaction is the formation of disk substructures \citep{CP2025, CP2025b, HK2026}. As shown in Figure~\ref{fig:spiralsnapshot}, the resulting disks exhibit non-axisymmetric spiral features. To delineate these structures more clearly, \cref{fig:primaryfit} presents the perturbed surface density, $\delta\Sigma/\langle\Sigma\rangle$,  at $t=8\kyr$, where $\langle\Sigma\rangle$ denotes the azimuthally averaged surface density and $\delta\Sigma \equiv \Sigma - \langle\Sigma\rangle$. It reveals a wide variety of substructures, including the primary spiral arm traced by dashed curves at large radii and secondary arms marked by dot-dashed curves at smaller radii. These secondary arms comprise two distinct families: weak, tightly-wound spiral density waves and more open logarithmic arms. In addition, a non-axisymmetric, crescent-shaped overdensity is visible in the upper-left region of each panel. The morphology and prominence of these features depend strongly on the mass infall rate, becoming progressively more pronounced with increasing $\dot{M}_{\rm in}$.  

\begin{figure*}[htb!]
    \epsscale{1.0}
    \plotone{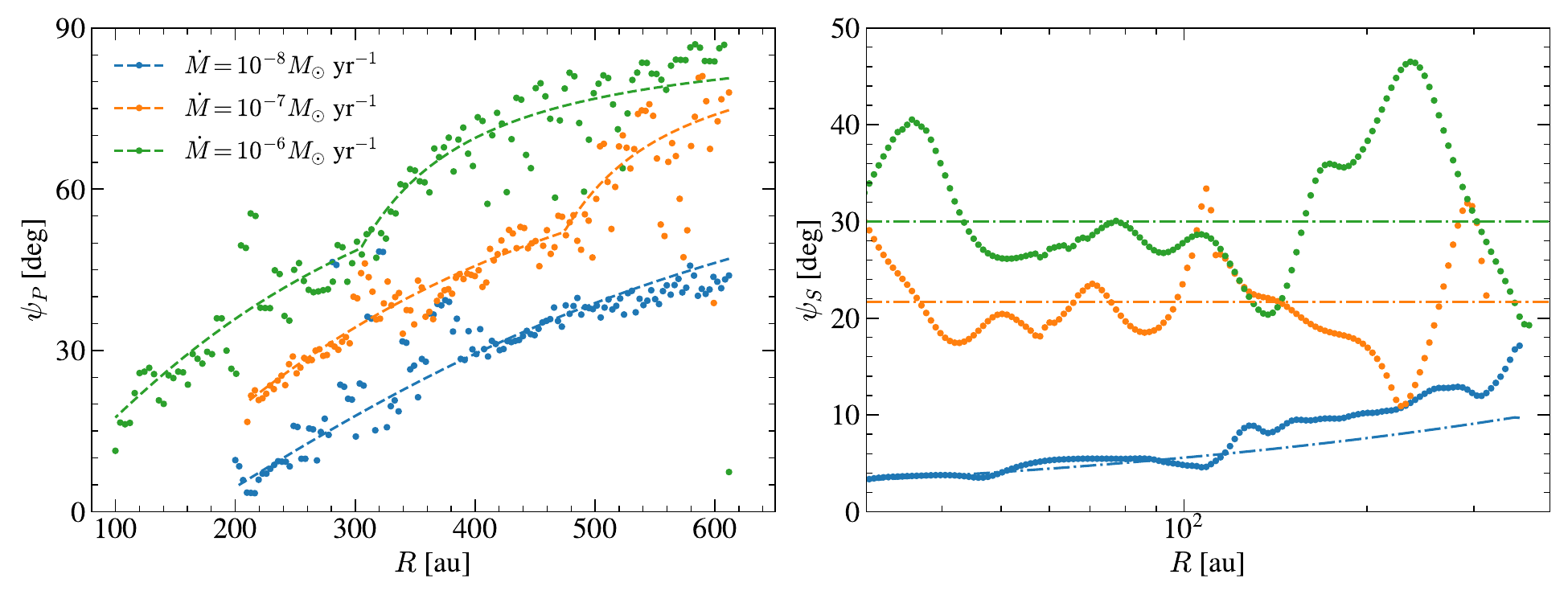}
    \caption{
    Pitch angles of the primary ($\psi_{P}$, \textit{left}) and secondary ($\psi_{S}$, \textit{right}) spiral arms measured at $t=8\kyr$ for models $\dot{M}_{\mathrm{in}}=10^{-8}\mdotyr$ (\textit{blue}), $10^{-7}\mdotyr$ (\textit{orange}), and $10^{-6}\mdotyr$ (\textit{green}). For the primary arm, the dashed curves show fits based on \cref{eq:primarylow} for the $\dot{M}_{\mathrm{in}}=10^{-8}\mdotyr$ model and \cref{eq:primaryhigh} for the $\dot{M}_{\mathrm{in}}=10^{-7}$ and $10^{-6}\mdotyr$ models. For the secondary arm, the dot-dashed lines plot the prediction of \cref{Eq:denwvapprox} for the $\dot{M}_{\mathrm{in}}=10^{-8}\mdotyr$ model, while constant pitch angles corresponding to  logarithmic spirals are adopted for the $\dot{M}_{\mathrm{in}}=10^{-7}$ and $10^{-6}\mdotyr$ models.}
    \label{fig:pitchangle}
\end{figure*}

The primary spiral arm originates in the outer disk from the interaction between the rotating disk gas and the infalling streamer, whose impact point remains nearly fixed in space during the infall phase. To quantify the geometry of the primary arm, we measure its pitch angle $\mathbf{\psi_{P}}$ as follows. We first determine the azimuthal location of the arm  $\phi_\text{peak}(R)$, as a function of radius. Starting from the outer radial boundary ($R=700\au$) near the streamer location ($\phi=0$), we identify the peak of the surface density perturbation $\delta\Sigma/\langle\Sigma\rangle$. We then trace the arm inward by iteratively locating the corresponding peak at smaller radii, constraining the azimuthal search range by the arm position identified at the previous radial step. Once the arm trajectory has been determined, the pitch angle is calculated from 
\begin{equation}
\frac{1}{\tan\psi(R)} \equiv  - R \frac{d\phi_\text{peak}(R)}{dR}\,,
\end{equation}
and the resulting radial profile is shown in the left panel of \cref{fig:pitchangle}.

In all models, the pitch angle of the primary arm decreases monotonically toward smaller radii, indicating that the arm becomes progressively more tightly wound as it propagates into the inner disk. For the model with $\dot{M}_{\mathrm{in}}=10^{-8}\mdotyr$, the radial variation of the pitch angle is well described by the linear function
\begin{equation}
    \tan\psi_{P}(R) = \eta R+\epsilon\,, \label{eq:primarylow}
\end{equation}
where $\eta\simeq 0.024$ and $\epsilon\simeq-0.40$ are the slope and the intercept of the fit, respectively.

In contrast, the pitch-angle profiles for the models with $\dot{M}_{\mathrm{in}}=10^{-7}\mdotyr$ and $10^{-6}\mdotyr$ are better described by a broken linear relation
\begin{equation}
    \tan\psi_{P}(R) = 
    \begin{cases}
        \eta_1(R-R_{\mathrm{tr}}) + \epsilon\,, &\text{for }R\leq R_{\mathrm{tr}}\,, \\
        \eta_2(R-R_{\mathrm{tr}}) + \epsilon\,, &\text{for }R >   R_{\mathrm{tr}}\,,
    \end{cases}
    \label{eq:primaryhigh}
\end{equation}
where $R_{\mathrm{tr}}$ denotes the transition radius at which the slope changes. For the  $\dot{M}_{\mathrm{in}}=10^{-7}\mdotyr$ model, the best-fit parameters are $\eta_1 \simeq 0.034, \eta_2\simeq 0.17, R_{\mathrm{tr}}\sim 474\au$, and  $\epsilon\simeq 1.28$. For the  $\dot{M}_{\mathrm{in}}=10^{-6}\mdotyr$ model, the corresponding values are $\eta_1 \simeq 0.041, \eta_2\simeq 0.16, R_{\mathrm{tr}}\simeq 305\au$, and  $\epsilon\simeq 1.15$. \cref{fig:primaryfit} compares the primary-arm trajectories measured from the simulations with those reconstructed from \cref{eq:primarylow,eq:primaryhigh}.

The broken-linear behavior of the primary-arm pitch angle for models with $\dot{M}_{\mathrm{in}}\geq 10^{-7}\mdotyr$ can be understood in terms of the ram pressure of the infalling streamer and the rotational shear of the disk. At large radii ($R>R_{\mathrm{tr}}$), the ram pressure of the infalling material is sufficiently strong that the flow remains largely unaffected by the disk rotation, allowing the primary arm to retain a nearly constant pitch-angle gradient. At smaller radii $(R<R_{\mathrm{tr}}$), however, the influence of disk rotation becomes increasingly important, causing the primary arm to be progressively sheared and thus altering its pitch-angle profile. This interpretation also explains why $R_{\mathrm{tr}}$ lies closer to the central star in the $\dot{M}_{\mathrm{in}}=10^{-6}\mdotyr$ model than in the $\dot{M}_{\mathrm{in}}=10^{-7}\mdotyr$ model: a higher mass infall rate corresponds to a larger ram pressure, enabling the infalling material to resist rotational shear down to smaller radii.

In addition to the primary spiral arm, all models exhibit a secondary spiral feature.  For the low-infall-rate model,  $\dot{M}_{\mathrm{in}}=10^{-8}\mdotyr$, we find that this structure is well described by a spiral density wave.

According to the linear WKB density-wave theory \citep{GT1979}, the wavefront of a non-self-gravitating spiral pattern is given by
\begin{equation}
    \phi_m(R) = \phi(R_p) \pm \int^R_{R_p}\frac{\sqrt{(\Omega(R')-\Omega_p)^2-\kappa^2/m^2}}{c_s(R')}dR'\,, \label{eq:lindenwave}
\end{equation}
where $R_p$ is the launching radius of the density wave, $\Omega_p$ is the pattern speed, and the sign specifies the propagation direction of the wave. Assuming a Keplerian disk ($\kappa\simeq\Omega$) and noting that the streamer remains at a nearly fixed azimuthal location throughout the simulation ($\Omega_p\simeq 0$), \cref{eq:lindenwave} reduces to
\begin{align}
    \phi_m(R) \sim \phi(R_p)\pm&\frac{2\sqrt{m^2-1}}{m(q_T-1)h_0} \notag \\ &\times\left[\left(\frac{R}{R_0}\right)^{(q_T-1)/2} - \left(\frac{R_p}{R_0}\right)^{(q_T-1)/2}\right]\,, \label{Eq:denwvapprox}
\end{align}
where $q_T$ is the radial temperature exponent and $h_0$ is the disk aspect ratio at $R_0$.

The dot-dashed line in the left panel of \cref{fig:primaryfit} compares the trajectory of the secondary spiral arm in the $\dot{M}_{\mathrm{in}}=10^{-8}\mdotyr$ simulation with that predicted by \cref{Eq:denwvapprox} with $m=2$. The semi-transparent line shows the identical trajectory shifted in phase by $180^\circ$.  Although the observed secondary arms broadly follow the expected shape of a density wave, they are systematically more loosely wound than the theoretical prediction, particularly in the outer disk. We attribute this discrepancy to the limitations of the WKB approximation, which becomes less accurate near the wave-launching region.

When the infall rate is sufficiently high, radial gradients in both the eccentricity and apoapsis position (\cref{fig:orbitalelements}) give rise to an additional logarithmic spiral arm.
The middle and right panels of \cref{fig:primaryfit} show that the spiral density waves identified in the low-infall-rate model are still present for $\dot{M}_{\mathrm{in}}\geq10^{-7}\mdotyr$. 
However, their signatures become much less prominent, overwhelmed by the stronger logarithmic spiral pattern generated by the eccentric disk structure.

We measure the trajectory of the logarithmic spiral arms using a procedure similar to that adopted for the primary arm. To isolate the large-scale spiral structure, we first apply Gaussian smoothing to the surface density field, thereby suppressing small-scale perturbations associated with the density waves. The resulting pitch angles, shown in the right panel of \cref{fig:pitchangle}, are $21.7^\circ$ and $30.0^\circ$ for the $\dot{M}_{\mathrm{in}}=10^{-7}\mdotyr$ and $10^{-6}\mdotyr$ models, respectively. The corresponding arm trajectories are overplotted in \cref{fig:primaryfit} as dot-dashed lines.

According to \citet{CP2025}, such spiral structures arise from orbital crowding caused by radial gradients in the disk eccentricity and apsidal orientation. The eccentric disk also develops a non-axisymmetric, crescent-shaped overdensity, visible in the upper-left region of each panel in \cref{fig:primaryfit}. Previous studies have shown that crescent-like structures can arise from vortices generated by Rossby-Wave instability \citep{RJ2012, LM2012, L2012, bae2015}, or from eccentric cavities driven by massive planets \citep{KD2006, RD2017, RK2023}. In our simulations, however, the crescent-shaped structure appears naturally from disk eccentricity excited by streamer accretion \citep{Deng2026}. The infalling material follows highly eccentric, nearly parabolic trajectories and transfers eccentricity to the disk gas. As a result, material tends to accumulate near the apoapsides of eccentric orbits, where the orbital velocity is lowest, producing the crescent-shaped overdensity.

\begin{figure}[htb!]
    \epsscale{1.15}
    \plotone{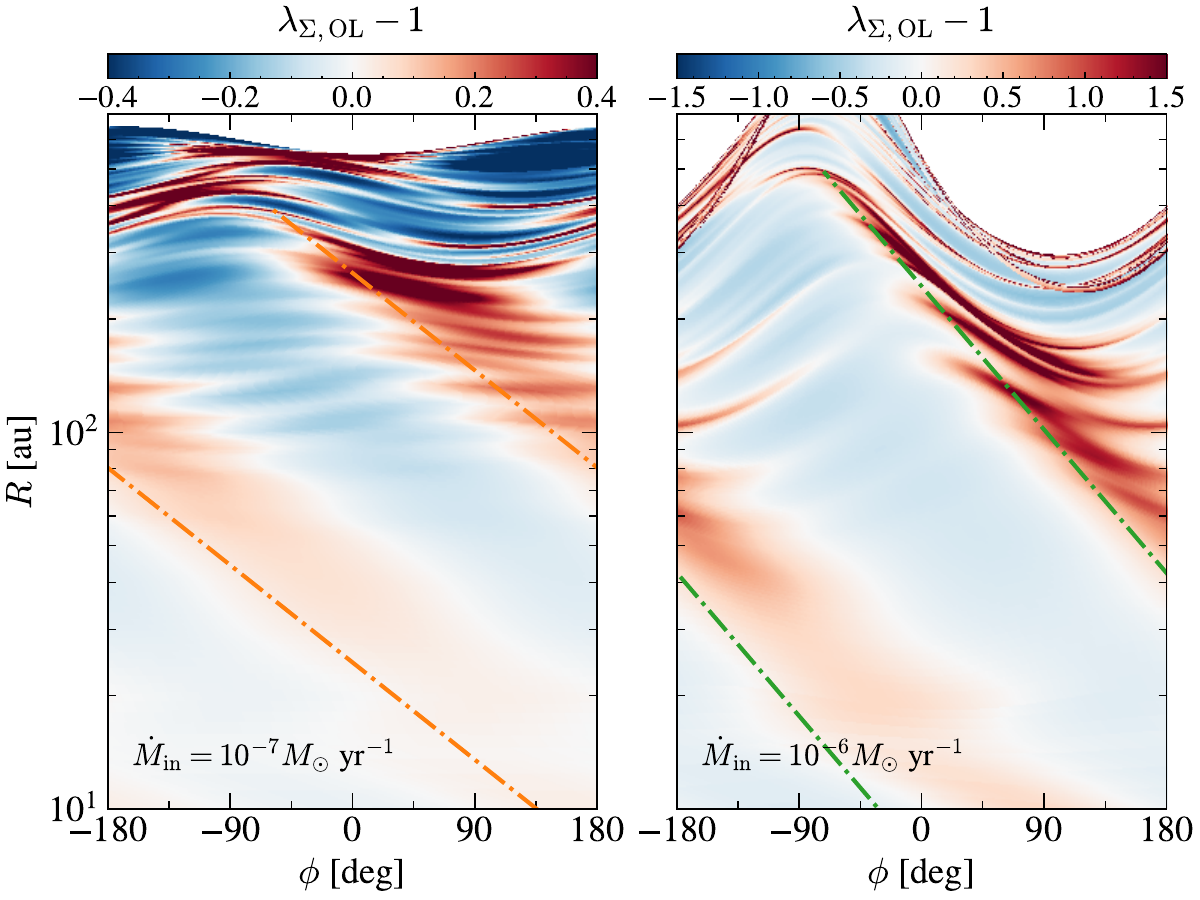}
    \caption{The surface density modulation factor $\lambda_{\Sigma}$ obtained from \cref{eq:OL_j} ($\lambda_{\Sigma,\mathrm{OL}}$) for $\dot{M}_{\mathrm{in}}=10^{-7}\mdotyr$ (\textit{left}) and $10^{-6}\mdotyr$ (\textit{right}) at $t=8\kyr$. The trajectories of the logarithmic spiral arms inferred from \cref{fig:pitchangle} are overplotted as dot-dashed curves. The logarithmic spiral arms and the locations of crescent-like structures closely follow the peaks of $\lambda_{\Sigma,\mathrm{OL}}$, supporting their origin in the disk eccentricity excited by the accretion streamer.}
    \label{fig:OL_surfdens}
\end{figure}

To test whether both the spiral and crescent-like structures can be explained by disk eccentricity, we again compare our simulation results with the analytic framework of \citet{OL2019}. In an eccentric disk, the surface density $\Sigma$ is modulated relative to that of a reference circular disk $\Sigma_o$, by the factor $\lambda_{\Sigma}\equiv 1/j$. \Cref{fig:OL_surfdens} presents $\lambda_{\Sigma}$ calculated from \cref{eq:OL_j}, together with the measured trajectories of the logarithmic spiral arms. In both the $\dot{M}_{\mathrm{in}}=10^{-7}\mdotyr$ and $10^{-6}\mdotyr$ models, the spiral arms closely follow the local maxima of $\lambda_{\Sigma,\mathrm{OL}}$. The crescent-like structures in the outer disk also coincide with regions where $\lambda_{\Sigma,\mathrm{OL}}$ is enhanced. This remarkable agreement indicates that both structures naturally follow from the disk eccentricity excited by the accretion streamer. Although \citet{CP2025} found that the resulting pattern is better described by a hyperbolic spiral, whereas our simulations produce logarithmic spirals, this difference is likely attributable to the different radial profiles of eccentricity and apsidal orientation resulting from the distinct infall prescriptions adopted in the two studies.

\subsection{Disk Angular Momentum} \label{subsec:angmom}

As the streamer material accretes onto the disk, it continuously supplies external angular momentum to the system. For our fiducial models, in which the infall region is located at $\cos\theta_0\in[0.5,0.6]$, the angular momentum injection rate $\dot{L}_{\mathrm{inj}}$ relative to the total disk angular momentum $L_{\mathrm{disk}}$ is 
\begin{equation}
\frac{\dot{L}_{\mathrm{inj}}}{L_{\mathrm{disk}}}= 68.8\left(\frac{\dot{M}_{\mathrm{in}}}{\msun\ \mathrm{yr}^{-1}}\right)\mathrm{yr}^{-1}\,. \label{eq:angmominjrate}
\end{equation}
Over the $8\kyr$ infall phase, therefore, the cumulative injected angular momentum amounts to $\sim 55\%$, $\sim 5.5\%$, and $\sim 0.55\%$ of the initial disk angular momentum for mass infall rates of $\dot{M}_{\mathrm{in}}=10^{-6}$, $10^{-7}$, and $10^{-8} \mdotyr$, respectively.

Because the orbital angular momentum of the infalling material in our fiducial model is inclined by $\sim 33^{\circ}$ with the disk rotation axis, the continuous injection of angular momentum can significantly modify the orientation of the disk angular momentum vector. In the high-infall-rate models, the streamer supplies a substantial fraction of the disk's initial angular momentum, suggesting that the disk may undergo a pronounced dynamical response. Previous studies have likewise highlighted the impact of late infall on disk orientation. For example, \citet{GF2021} presented observational evidence that streamer accretion can induce a misalignment between the inner and outer regions of a disk,  while \citet{KD2021} demonstrated, using a cloudlet infall model, that late infall can generate significant disk misalignment. Motivated by these findings, we investigate how streamer accretion modifies the disk's angular momentum distribution and whether it can induce warping or even produce a misaligned disk configuration.

\begin{figure*}[htb!]
    \epsscale{1.0}
    \plotone{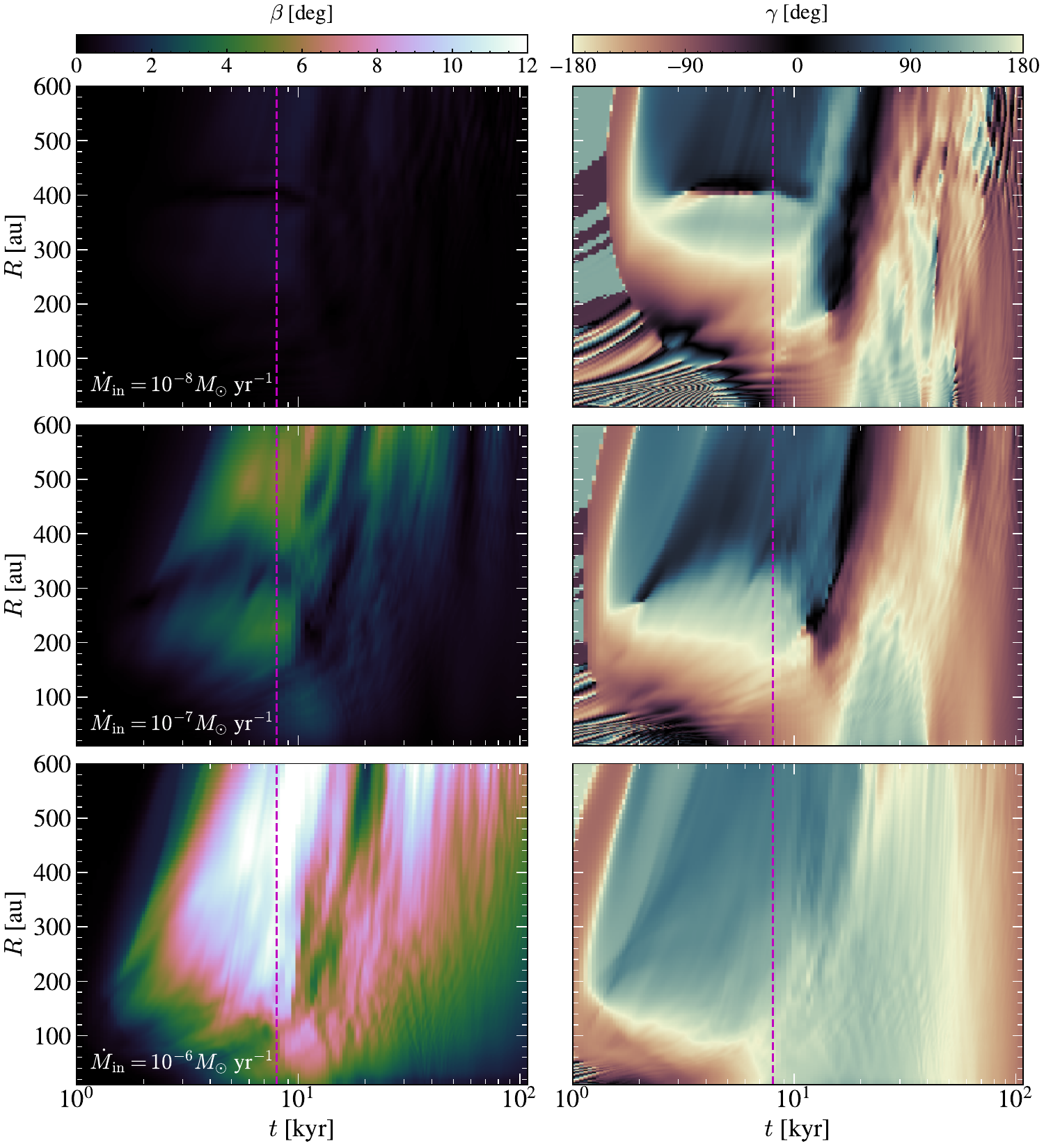}
    \caption{Temporal and radial evolution of the tilt angle $\beta$ (\textit{left}) and twist angle $\gamma$ (\textit{right}) for models with  $\dot{M}_{\mathrm{in}}=10^{-8}\mdotyr$ (\textit{top}), $10^{-7}\mdotyr$ (\textit{top}) and $10^{-6}\mdotyr$ (\textit{bottom}).}
    \label{fig:warpparam}
\end{figure*}

To quantify the effect of streamer accretion on the disk angular momentum, we defined two orientation angles
\begin{align}
    \beta &= \arccos (L_z /|\mathbf{L}|)\,,\\
    \gamma &= \arctan(L_y/L_x)\,,
\end{align}
where $L_x, L_y$, and $L_z$ are the  Cartesian components of the angular momentum vector $\mathbf{{L}}=\int \rho \mathbf{r}\times\mathbf{v}\,dz\,d\phi$. Here, $\beta$ and $\gamma$ represent the tilt and twist angles of the disk, respectively.

\cref{fig:warpparam} plots the temporal and radial distributions of these quantities for the fiducial models, demonstrating that streamer accretion with an infall rate of $\dot{M}_{\mathrm{in}}=10^{-8}\mdotyr$ has only a minor effect on the disk angular momentum orientation. In contrast, models with $\dot{M}_{\mathrm{in}}=10^{-7}\mdotyr$ and $10^{-6}\mdotyr$ develop significant warps, with the outer disk reaching tilt angles of $\sim 5^\circ$, and $\sim 12^\circ$, respectively. Unlike the eccentricity profiles shown in Figure~\ref{fig:orbitalelements}, the tilt angle increases monotonically with disk radius. This behavior arises because the inner disk can reorient itself through angular momentum exchange with the outer disk, a process governed by a physical mechanism fundamentally different from that responsible for eccentricity growth \citep{FI2026}.

Moreover, we find a modest misalignment between the inner and outer disks in the $\dot{M}_{\mathrm{in}}=10^{-7}$ and $10^{-6}\mdotyr$ models. This is characterized by a difference in the twist angle of $\sim 180^\circ$, accompanied by differences in the tilt angle of only $\sim 4^\circ$ and $\sim 6^\circ$, respectively. However, these mild disk misalignments gradually weaken after the infall ceases. In the $\dot{M}_{\mathrm{in}}=10^{-6}\mdotyr$ model, the disk misalignment nearly disappears by the end of the infall phase as angular momentum is redistributed from the outer disk to the inner disk, promoting misalignment of the disk as a whole.

\subsection{Accretion Rate} \label{subsec:accrate}

\begin{figure}[htb!]
    \epsscale{1.15}
    \plotone{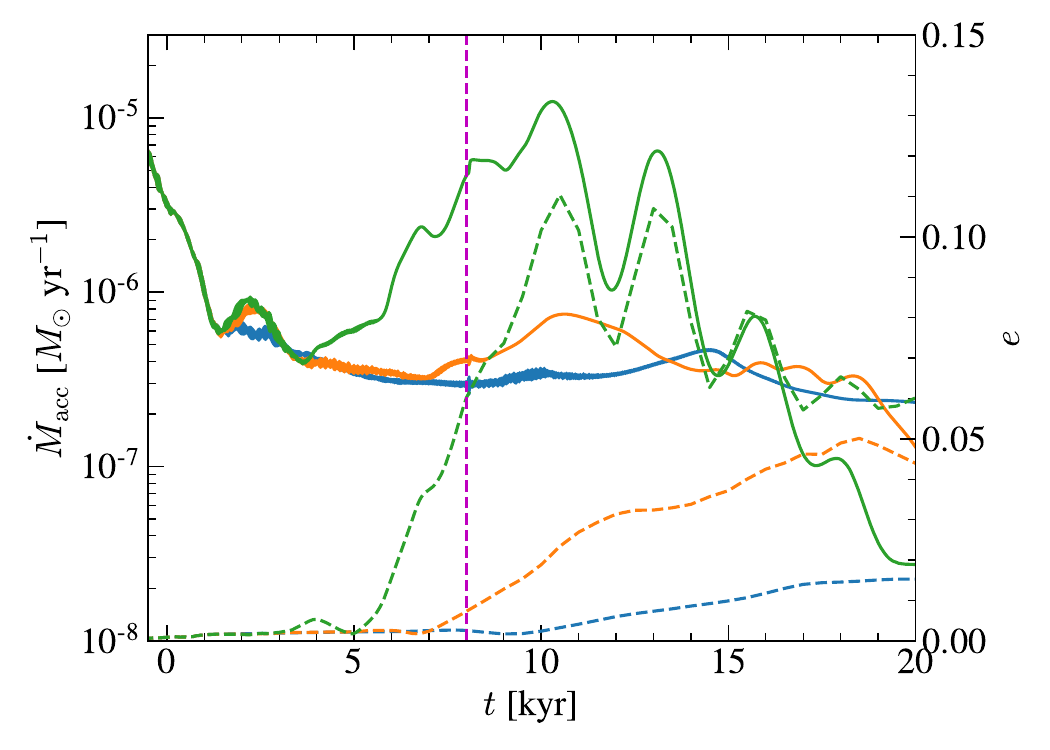}
    \caption{Temporal evolution of the stellar accretion (\textit{solid lines}) and the mean disk eccentricity for $a\leq10\au$ (\textit{dashed lines}) for models with $\dot{M}_{\mathrm{in}}=10^{-8}\mdotyr$ (\textit{blue}), $10^{-7}\mdotyr$ (\textit{orange}), and $10^{-6}\mdotyr$ (\textit{green}). The vertical dashed line indicates the phase where the infall is ceased, similar to \cref{fig:orbitalelements} and \ref{fig:warpparam}. The accretion rate in the  $\dot{M}_{\mathrm{in}}=10^{-6}\mdotyr$ model is enhanced by a factor of $\sim30$ relative to the low-infall rate models. In addition, it exhibits quasi-periodic variations with a period of $\sim 3\kyr$, which closely track the evolution of the inner-disk eccentricity.}
    \label{fig:accretionrate}
\end{figure}

Various recent observations suggest a possible connection between disk accretion and streamer infall \citep{HG2024, WB2024}. In this section, we investigate how streamer accretion affects disk accretion.

\cref{fig:accretionrate} plots the temporal evolution of the accretion rate $\dot{M}_{\mathrm{acc}}$ onto the central star, for $\dot{M}_{\mathrm{in}}=10^{-8}\mdotyr$, $10^{-7}\mdotyr$, and $10^{-6}\mdotyr$. At early times ($t\lesssim1.5\kyr$), the influence of the streamer has not yet propagated through the disk to the innermost regions. During this phase, the measured accretion rate is driven primarily by initial disk relaxation and viscous evolution. For the $\dot{M}_{\mathrm{in}}=10^{-8}\mdotyr$ model, we observe only a modest increase in the accretion rate following the cessation of infall, reaching a maximum at $t\sim 15$ kyr. In contrast, models with $\dot{M}_{\mathrm{in}}\geq 10^{-7}\mdotyr$ exhibit two distinct episodes of enhanced accretion.  The first enhancement occurs at $t\sim1.5\kyr$, when the accretion rate increases by a factor of $\sim 2$, consistent with the result of \citet{CP2025}. This early enhancement likely arises from collisions between infalling streamer material and rotating disk gas along the primary spiral arm, which promote angular momentum transport and thus enhance mass accretion. However, this effect is transient and persists only until $t\sim 10\kyr$. In particular, the $\dot{M}_{\mathrm{in}}=10^{-7}\mdotyr$ model shows little difference from the $\dot{M}_{\mathrm{in}}=10^{-8}\mdotyr$ model over an extended period, despite the much more prominent spiral structures discussed in the previous section. This suggests that strong spiral structures alone are insufficient to sustain enhanced accretion over long timescales.

A second episode of enhanced accretion occurs in models with $\dot{M}_{\mathrm{in}}\ge10^{-7}\mdotyr$. This enhancement begins at $t\sim7\kyr$ and $\sim 3.5\kyr$ for $\dot{M}_{\mathrm{in}}=10^{-7}\mdotyr$ and $10^{-6}\mdotyr$, respectively. Notably, the accretion rate continues to increase even after infall ceases, reaching a maximum at $t\sim 10.5$ kyr in both models. The peak accretion rate exceeds that of the corresponding no-infall phase by factors of $\sim 2$ and $\sim 30$ for the $\dot{M}_{\mathrm{in}}=10^{-7}\mdotyr$ and $10^{-6}\mdotyr$ models, respectively. In addition, the $\dot{M}_{\mathrm{in}}=10^{-6}\mdotyr$ model exhibits quasi-periodic oscillations in the accretion rate with a period of $\sim 3\kyr$. These oscillations persist even after infall ceases. 

This second episode of enhanced accretion, including the quasi-periodic oscillation observed in the  $\dot{M}_{\mathrm{in}}=10^{-6}\mdotyr$ model, can be understood as a consequence of the eccentricity excited by streamer accretion \citep{CP2025}. As the eccentricity of the inner disk grows, a larger fraction of disk gas reaches smaller stellocentric distances at pericenter, thereby enhancing accretion onto the central star. Furthermore, because eccentricity propagates through the disk in a wave-like manner and undergoes oscillatory evolution \citep{OB2014, OL2019}, the accretion rate is expected to exhibit corresponding temporal variations. The dashed curves in \cref{fig:accretionrate} show the eccentricity of the gas near the inner radial boundary for comparison with the accretion rate. We find a strong correlation between the two quantities, both in the large accretion outburst occurring at $t\sim5\kyr$ and the quasi-periodic oscillations with a period of $\sim 3\kyr$ observed in the $\dot{M}_{\mathrm{in}}=10^{-6}\mdotyr$ model. 

\section{Synthetic Observations} \label{sec:synobs}

We now turn to synthetic observations to explore the observational signatures of hydrodynamic phenomena discussed in Section \ref{sec:hydro}. To do so, we use RADMC-3D \citep{du2012} to generate synthetic molecular-line channel maps. We do not perform full Monte Carlo radiative transfer calculations to determine the dust temperature distribution, since the adopted initial temperature profile is already constrained by observations \citep{RF2020}. Nevertheless, test calculations using dust temperature distributions from Monte Carlo radiative transfer simulations yielded synthetic observations nearly indistinguishable from those presented here, indicating that our results are insensitive to the adopted temperature treatment. 

For this study, we focus on the $^{13}$CO $J=2\!-\!1$ transition, which has a rest frequency of $\nu_0=220.399$ GHz \citep{sv2005}. This transition has been widely used to trace gas kinematics in PPDs and \citet{SD2024} used it to probe kinematic perturbations in the AB Aur disk. We assume a $^{12}$CO-to-H$_2$ abundance ratio of $1\times 10^{-4}$ and a carbon isotope ratio of $[^{12}\mathrm{C}]/[^{13}\mathrm{C}]= 77$ \citep{wr1994}, from which $^{13}$CO  abundance is derived. Using these assumptions, we construct the three-dimensional $^{13}$CO number-density cube, along with the corresponding velocity cubes for the three velocity components, obtained from the hydrodynamic simulations. 

Our observational setup is based on recent ALMA observations of AB Aur presented by \citet{SD2024}. We adopt a disk inclination of $20^\circ$ and a distance of $155.9$ pc. The synthetic line cubes span a total velocity range of $12\kms$ and are divided into 300 spectral channels, corresponding to a velocity resolution of $0.04\kms$. 

After generating the synthetic line cubes from RADMC-3D, we use CASA \citep{CASA2022} to simulate realistic ALMA observations. We first employ the \textsc{simobserve} task to generate model visibilities using the antenna configurations adopted by \citet{SD2024}. Specifically, we simulate a $16308$ s integration in the ALMA C8.6 configuration and combine it with a $4392$ s integration in the C8.3 configuration, matching the observational setup of \citep{SD2024}. This procedure incorporates beam-convolution effects directly at the visibility level. We then use the \textsc{simulator.corrupt} task to inject thermal noise into the synthetic visibilities. We find that applying visibility noise with a level of $\sim 3$ Jy produces a channel rms noise level of $\sim 2$ mJy beam$^{-1}$, comparable to that achieved in the observations of \citet{SD2024}.

\subsection{Channel Maps} \label{subsec:channel}

\begin{figure*}[htb!]
    \epsscale{1.0}
    \plotone{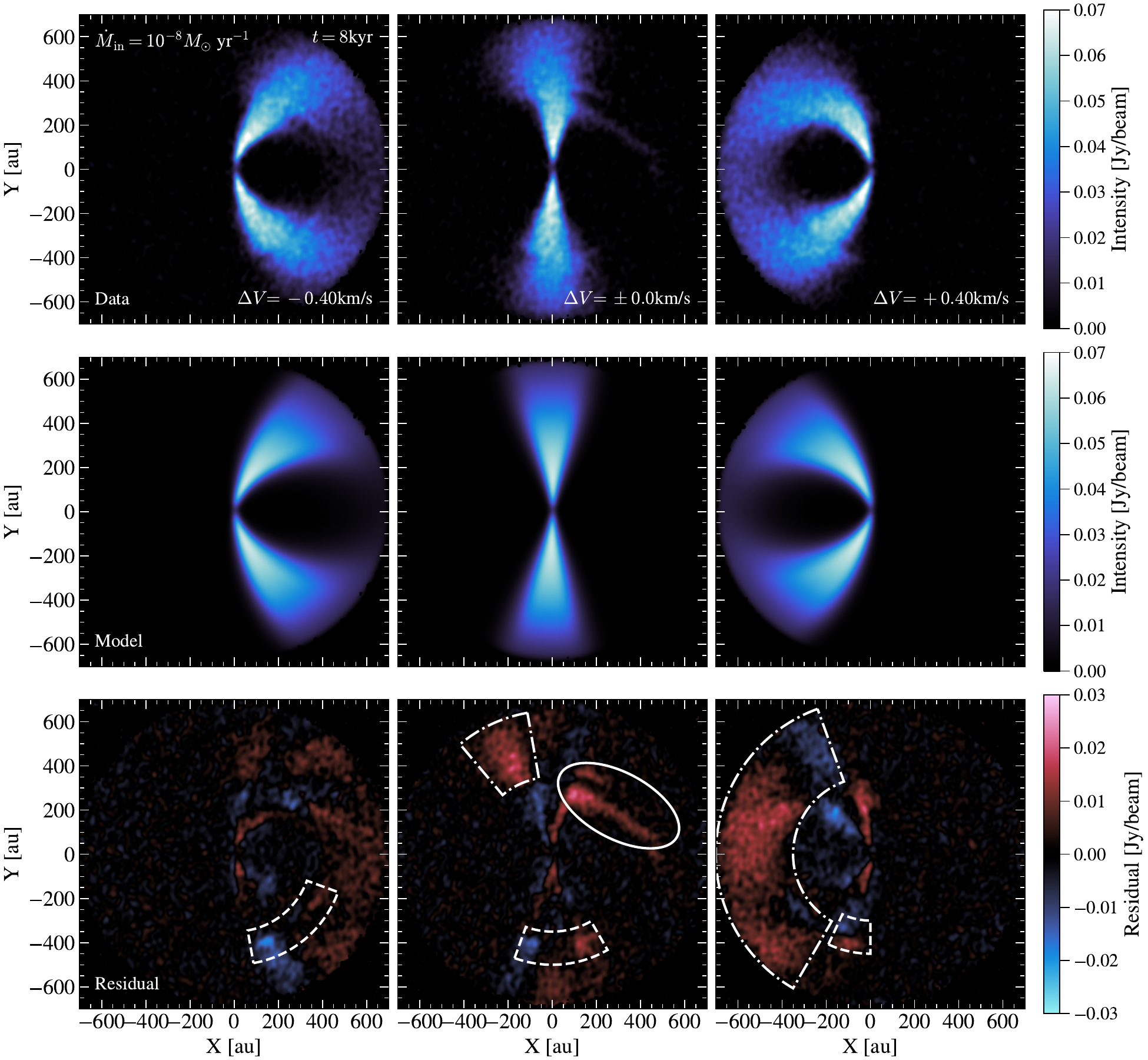}
    \caption{$^{13}$CO channel maps for the $\dot{M}_{\mathrm{in}}=10^{-8}\mdotyr$ model at $t=8\kyr$. The top, middle, and bottom panels show the synthetic observations, the best-fitting Keplerian model, and the residuals obtained by subtracting the model from the synthetic data, respectively. The residual maps reveal several signatures of streamer accretion, including the primary spiral arm (\textit{solid}), the crescent-shaped structure (\textit{dashed}), and the combined effects of the eccentric disk and disk warp (\textit{dot-dashed}).}
    \label{fig:channel_1}
\end{figure*}

We first present the channel maps for $\dot{M}_{\mathrm{in}}=10^{-8}\mdotyr$. \cref{fig:channel_1} shows  snapshots of the velocity channel maps taken at $t=8$ kyr. The top panels display the synthetic observations, the middle panels show the best-fitting Keplerian models, and the bottom panels present the residuals obtained by subtracting the model from the synthetic data.

The model data cube is constructed using the tool \textsc{discminer}\footnote{https://github.com/andizq/discminer/} \citep{IT2021}.
This tool employs Markov Chain Monte Carlo (MCMC) techniques to fit the line-intensity distributions, rotation velocity field, and emission surface geometry of the observed channel maps, thereby producing a Keplerian model data cube. We adopt the following rotational velocity profile:
\begin{equation}
    v_\phi = v_{\mathrm{sys}} + \sqrt{\frac{GM_*R^2}{(R^2+z_{\mathrm{em}}^2)^{3/2}}}  \,, 
\end{equation}
where $v_{\mathrm{sys}}$ is the systematic velocity of the system and $z_{\mathrm{em}}$ is the height of the emission surface. We parameterize the emission surface as 
\begin{align}
    z_{\mathrm{em}} (R) &= z_{\mathrm{em},0}\left(\frac{R}{R_0}\right)^{q_{\mathrm{em},1}}\exp\left[-\left(\frac{R}{R_b}\right)^{q_{\mathrm{em},2}}\right]\,,
\end{align}
where $z_{\mathrm{em},0}$ is the emission-surface height at the reference radius $R_0$, $q_{\mathrm{em},1}$ is the radial power-law index, $q_{\mathrm{em},2}$ represents the tapering index of the emission surface, and $R_b$ is the characteristic tapering radius for the emission surface.

In the channel maps, the most prominent feature induced by streamer accretion is the primary spiral arm. We find that its signature is most clearly visible near the systemic velocity ($\Delta V \approx 0\kms$), where it appears at $Y\sim0$--$400\au$ in \cref{fig:channel_1}. In addition, we identify velocity kinks on the opposite side of the disk ($Y<0\au$) in several velocity channels. Based on their locations relative to the hydrodynamic structures discussed in Section \ref{subsec:substr}, these features are likely associated with the crescent-shaped overdensity and the secondary spiral arms.

We also detect a pronounced asymmetry between the upper and lower lobes of the emission, a characteristic signature of a warped disk \citep{YA2022}. In particular, positive residuals are present at $X > 0\au$ and $Y \gtrsim 300\au$ across multiple velocity channels. These residuals are likely produced by the disk warp and are consistent with the tilt and twist structure shown in \cref{fig:warpparam}, which exhibits a twist-angle difference of  $\sim 45^\circ$ between the inner and outer regions.

\begin{figure*}[htb!]
    \epsscale{1.0}
    \plotone{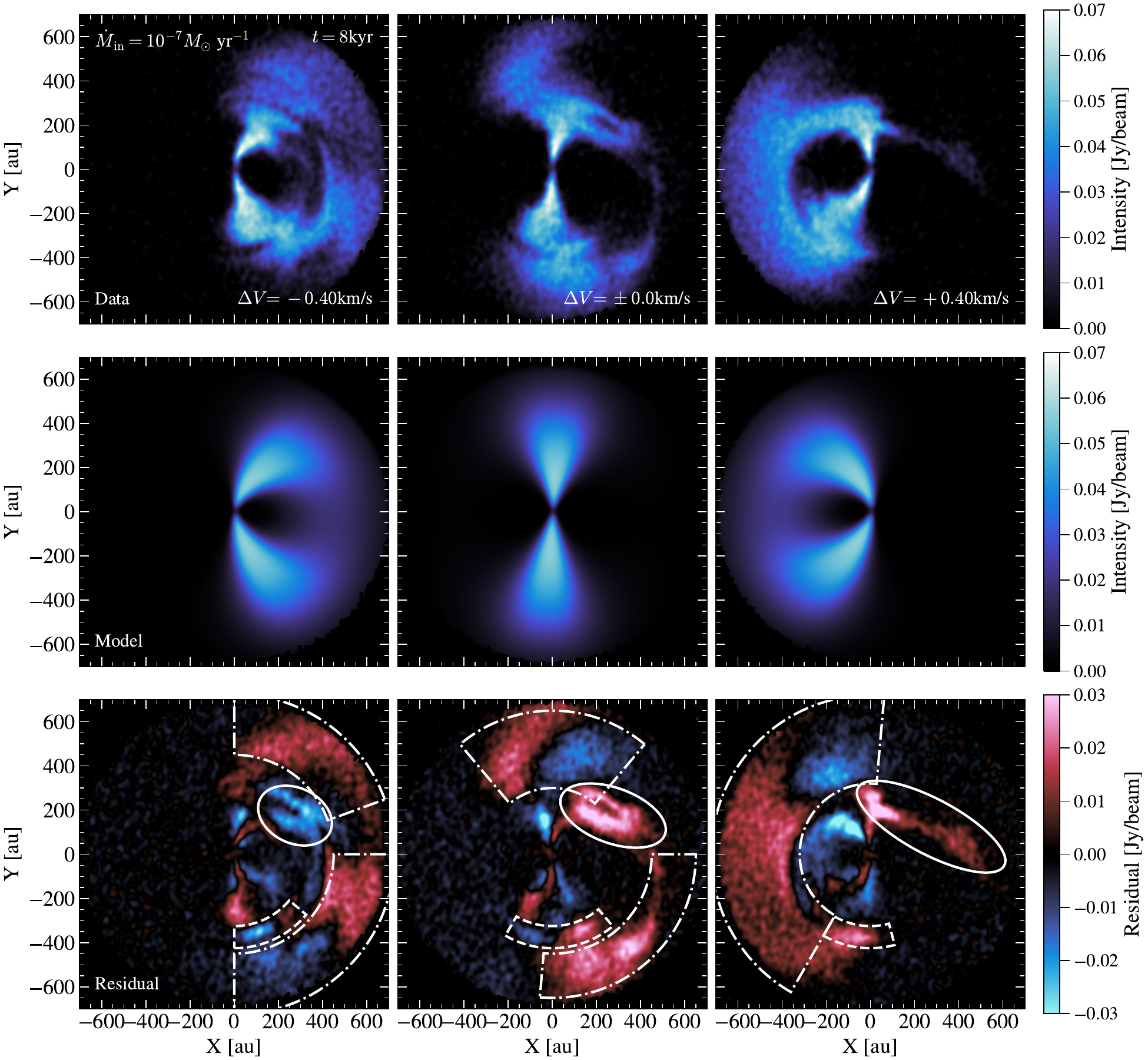}
    \caption{Same as \cref{fig:channel_1} but for the $\dot{M}_{\mathrm{in}}=10^{-7}\mdotyr$ model. The streamer-induced signatures are more prominent and span a broader velocity range than in the $\dot{M}_{\mathrm{in}}=10^{-8}\mdotyr$ model. The positive residuals in the outer disk (\textit{dot-dashed}) are also visible in the lower lobe, suggesting a stronger disk warp. Residuals are likewise present in the inner disk ($R<200\au$); however, these features should be interpreted with caution, as this may arise from fitting the data cube with a Keplerian circular-disk model.}
    \label{fig:channel_2}
\end{figure*}

\begin{figure*}[htb!]
    \epsscale{1.0}
    \plotone{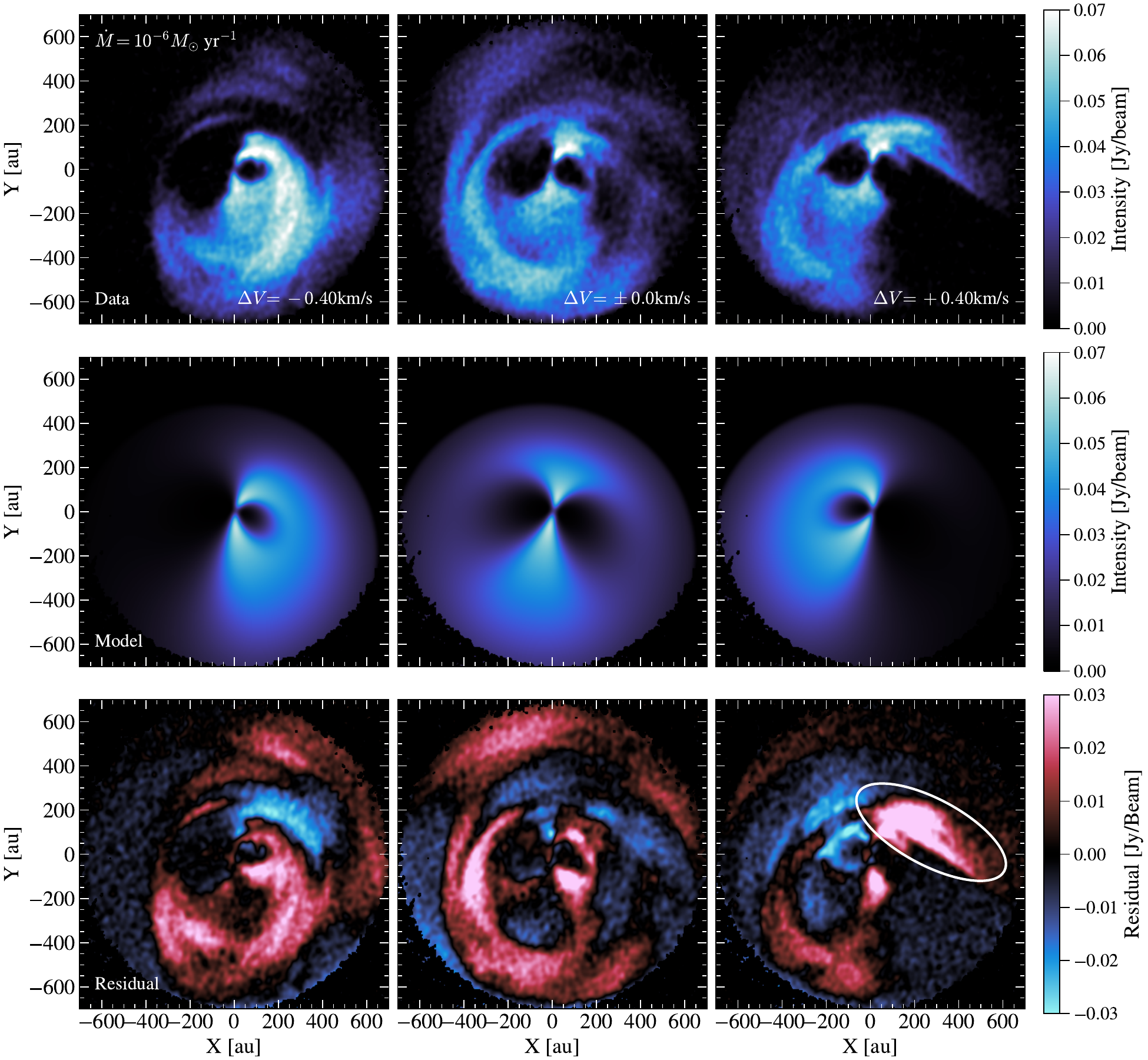}
    \caption{Same as Figures \ref{fig:channel_1} and \ref{fig:channel_2}, but for the $\dot{M}_{\mathrm{in}}=10^{-6}\mdotyr$ model. The residuals are dominated by the eccentric disk and disk warp, which produce prominent arc-like structures and significantly distort the disk's kinematic position angle.}
    \label{fig:channel_4}
\end{figure*}

As the mass infall rate increases, the streamer's dynamical impact on the disk grows stronger, making the associated observational signatures more prominent. This trend is evident in \cref{fig:channel_2}, which is analogous to \cref{fig:channel_1} but for the $\dot{M}_{\mathrm{in}}=10^{-7}\mdotyr$ model. The signature of the primary spiral arm  spans a broader velocity range, while the velocity kink on the opposite side of the disk becomes easier to identify. In addition, the warped geometry of the outer disk is more clearly visible through systematic shifts in the residual emission pattern. The warp is sufficiently pronounced that its signature can also be traced on the opposite side of the disk. Finally, the eccentric disk induced by streamer accretion produces arc-like residual structures in the outer disk, providing an additional observational signature of the perturbed disk geometry.

When the infall rate reaches $\dot{M}_{\mathrm{in}}=10^{-6}\mdotyr$, the streamer excites a sufficiently large disk eccentricity that the observed kinematics become dominated by eccentric motions. As a result, arc-like residual structures appear across multiple velocity channels, and the kinematic position angle of the disk becomes noticeably distorted. As shown in \cref{fig:channel_4}, this distortion shifts the kinematic signature of the primary spiral arm toward blueshifted velocities. Consequently, the positive residual associated with the primary spiral arms, analogous to that seen in Figures \ref{fig:channel_1} and \ref{fig:channel_2}, is detected only in the $\Delta V = +0.4\kms$ channel.

Although residuals are also present in the inner disk ($R<200\au$), their interpretation requires caution. At $t=8\kyr$ in the $\dot{M}_{\mathrm{in}}=10^{-7}\mdotyr$ model, the disk has already developed significant eccentricity and warp, making the assumption of a Keplerian circular disk increasingly inappropriate. As a result, the model data cubes generated by \textsc{DISCMINER} exhibit broader emission lobes than those shown in \cref{fig:channel_1}. This broadening arises because \textsc{DISCMINER} attempts to reproduce the velocity shifts induced by the eccentric disk and disk warp within the framework of a circular-disk model. Consequently, the mismatch between the model and underlying disk geometry can produce artificial residuals in the inner disk ($R \leq 200\au$).

\begin{figure*}[htb!]
    \epsscale{1.0}
    \plotone{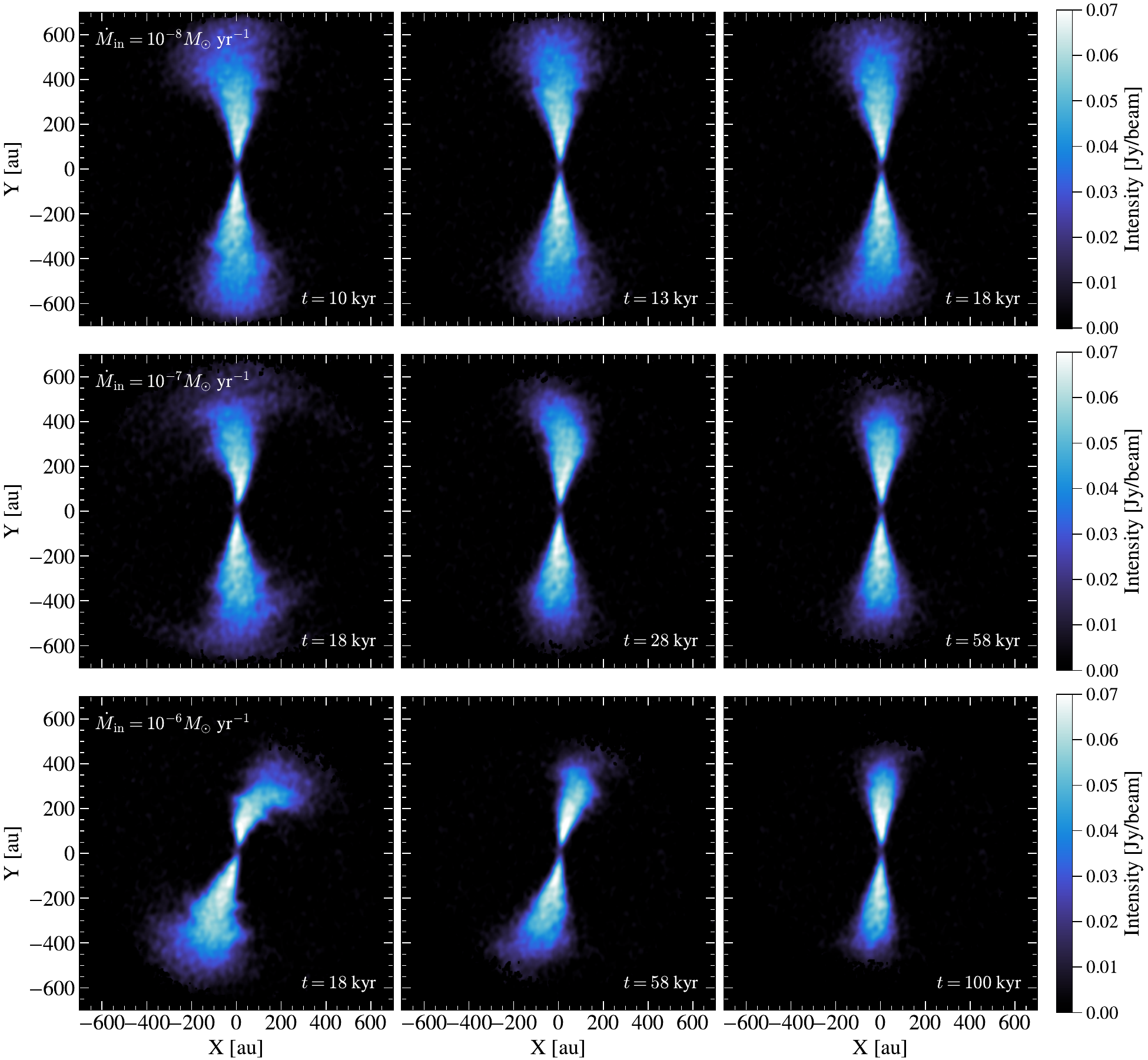}
    \caption{Snapshots of the $^{13}$CO channel maps at the systematic velocity for models with $\dot{M}_{\mathrm{in}}=10^{-8}\mdotyr$, (\textit{top}), $10^{-7}\mdotyr$ (\textit{middle}), and $10^{-6}\mdotyr$ (\textit{bottom}) after the cessation of infall. Streamer-induced kinematic signatures decay gradually, but their survival times increase with the mass infall rate.}
    \label{fig:channel_3}
\end{figure*}

We now investigate the longevity of these signatures after infall ceases. \cref{fig:channel_3} displays snapshots of the line-center channel maps at different epochs for three infall-rate models. Although the streamer-induced features gradually weaken, they persist much longer at higher infall rates. For $\dot{M}_{\mathrm{in}}=10^{-8}\mdotyr$, the disk returns to an approximately Keplerian state within $\sim10\kyr$. In contrast, the corresponding relaxation timescale increases to $\sim 50\kyr$ for the $\dot{M}_{\mathrm{in}}=10^{-7}\mdotyr$ model. For $\dot{M}_{\mathrm{in}}=10^{-6}\mdotyr$, the streamer-induced perturbations remain visible for nearly $100\kyr$. In particular, the kinematic position angle of the disk is significantly displaced from its initial value and gradually relaxes as the disk eccentricity decays (\cref{fig:orbitalelements}). We also find evidence for a long-lived eccentric disk in the form of asymmetric emission lobes, with the upper lobe appearing noticeably smaller than the lower lobe in the $\dot{M}_{\mathrm{in}}=10^{-6}\mdotyr$ model. 

\subsection{Moment Map} \label{subsec:moment}

\begin{figure*}[htb!]
    \epsscale{1.0}
    \plotone{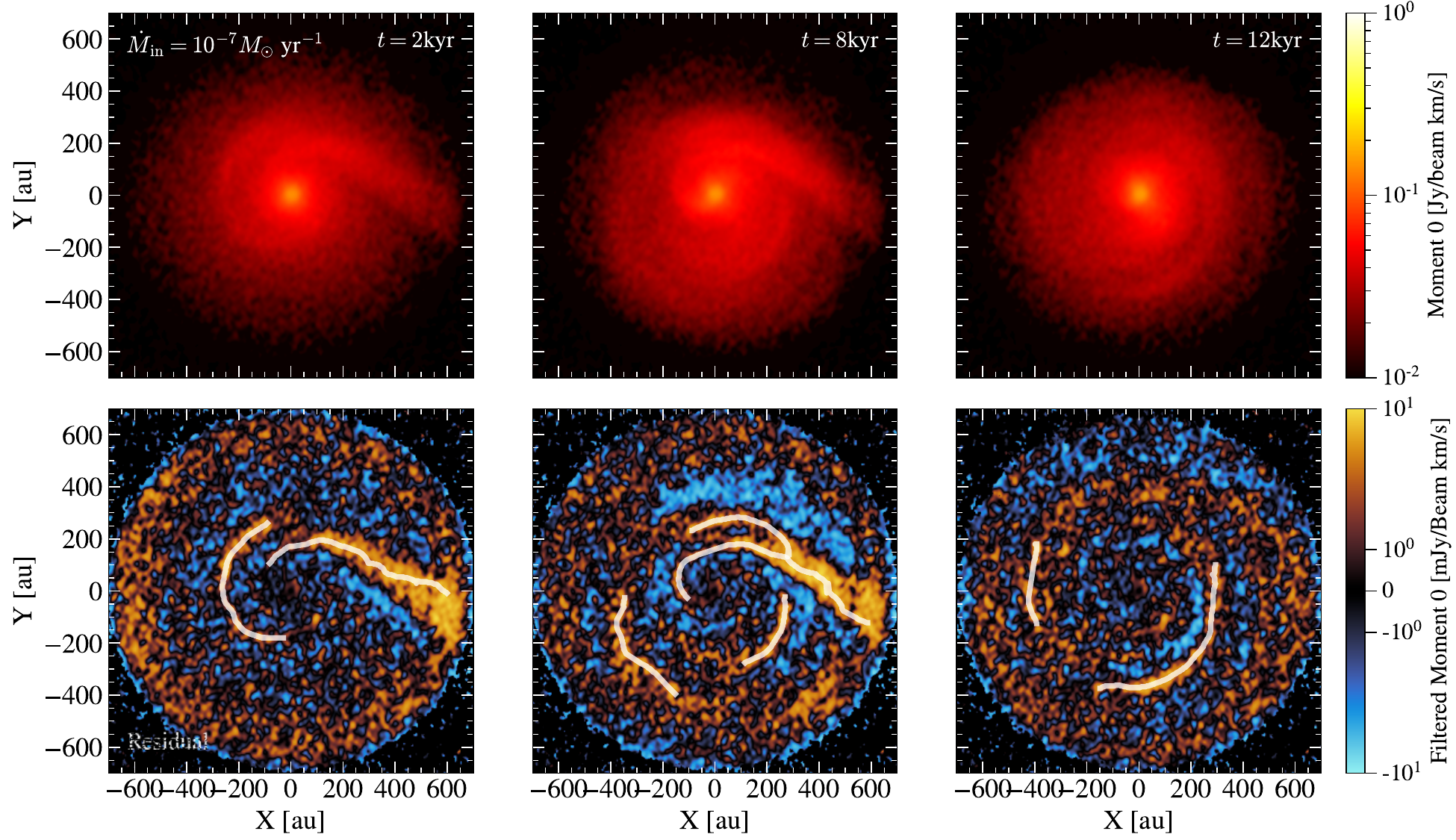}
    \caption{Moment-0 maps (\textit{top}) and filtered moment-0 maps (\textit{bottom}) derived from the synthetic $^{13}$CO observations of the $\dot{M}_{\mathrm{in}}=10^{-7}\mdotyr$ at $t=2$, $8$ and $12\kyr$. Spiral arms identified in the filtered moment-0 maps are overplotted as solid white curves.}
    \label{fig:moment0}
\end{figure*}

We first examine snapshots of the integrated-intensity (moment-0) maps for the $\dot{M}_{\mathrm{in}}=10^{-7}\mdotyr$ model during ($t=2, 8\kyr$) and after ($t=12\kyr$) the infall phase, as shown in \cref{fig:moment0}. While moment-map residuals are commonly constructed by subtracting a Keplerian model from the observed moment map, we instead adopt the ``Filtered moment" technique introduced by \citet{SD2024} and subsequently applied by \citet{CP2025b}.

In this method, the image is smoothed using a radially expanding kernel
\begin{equation}
    w(r) = w_0\left(\frac{r}{r_0}\right)^\zeta\,, \label{eq:kernel}
\end{equation}
before subtracting the smoothed image from the original. This procedure effectively removes large-scale structures, such as those associated with disk eccentricity and warping, while preserving small-scale features including spirals and other disk substructures. Throughout this work, we adopt $w_0=5$ pixels and $\zeta=0.5$. The resulting filtered moment-0 maps are plotted in the bottom panels of \cref{fig:moment0}. 

To identify spiral arms, we first smooth the filtered moment-0 maps with a Gaussian kernel to suppress noise. We then apply a Frangi ridge filter implemented in the \textsc{skimage-image} Python package to identify candidate ridge structures. By selecting ridge pixels with confidence levels above 95\%, we can trace the spiral arms in the filtered moment-0 maps. The resulting spiral-arm trajectories are overplotted as solid white curves in the bottom panels of Figure~\ref{fig:moment0}.

\begin{figure*}[htb!]
    \epsscale{1.0}
    \plotone{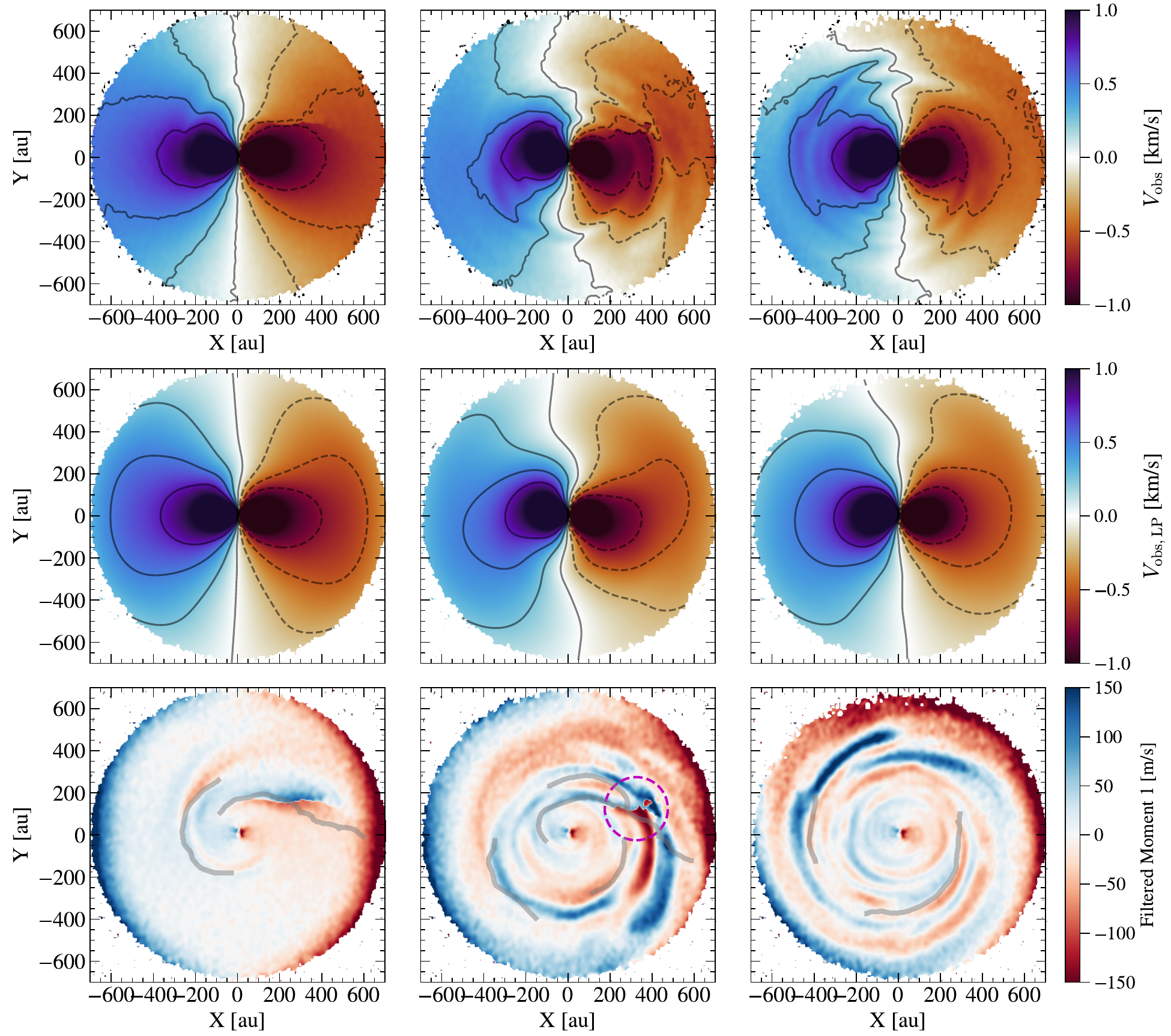}
    \caption{Moment-1 maps (\textit{top}), smoothed moment-1 maps obtained using the kernel defined in \cref{eq:kernel} (\textit{middle}), and filtered moment-1 maps (\textit{bottom}) for the $\dot{M}_{\mathrm{in}}=10^{-7}\mdotyr$ at $t=2, 4$, and $12\kyr$. The spiral arms identified from the filtered moment-0 maps and the Doppler-flip locations are overplotted in the bottom panels as solid gray curves and a dashed magenta circle, respectively. The smoothed moment-1 maps primarily highlight the large-scale disk warp induced by streamer accretion, while the filtered moment-1 maps reveal velocity perturbations associated with streamer-induced disk substructures.}
    \label{fig:moment1}
\end{figure*}

Next, we compare the spiral arms identified in the filtered moment-0 maps to the intensity-weighted line-of-sight velocity (moment-1) maps presented in \cref{fig:moment1}. The middle row displays moment-1 maps smoothed using the kernel defined in \cref{eq:kernel}. These maps emphasize the large-scale kinematic distortions induced by streamer accretion. In particular, the contours corresponding to line-of-sight velocities between $-0.75\kms$ and $0.75\kms$ exhibit noticeable twist between the inner and outer disk regions, a characteristic signature of a warped disk \citep{YA2022, SD2024}.

We next examine the moment-1 residuals shown in the bottom row and compare them with the overplotted spiral arms. Previous studies have shown that spiral arms driven by gravitational instability are associated primarily with radially convergent flows \citep{HD2020, LL2021}, whereas planet-driven spirals tend to produce predominantly radially divergent perturbations \citep{GR2001, R2002}. Consistent with the findings of \citet{CP2025b}, we find that streamer-induced spiral structures generate both convergent and divergent velocity perturbations. By comparing the residual maps with snapshots from our hydrodynamic simulations, we infer that the radially divergent perturbations originate from disk material accelerated by the ram pressure of the infalling gas, while the spiral arms discussed in Section \ref{subsec:substr} are associated primarily with convergent motions. 

Although the spiral arms identified in the filtered moment-0 maps correlate well with the moment-1 residuals, we also detect several kinematic features in the moment-1 residuals that have no clear counterparts in the moment-0 maps. This demonstrates that molecular-line kinematics provide a more sensitive diagnostic of past infall events than intensity maps alone.

Finally, we identify a Doppler flip, a localized reversal in the sign of the velocity residuals, during the active infall phase, indicated by the dashed line in \cref{fig:moment1}. Although Doppler flips are commonly interpreted as signatures of embedded planets \citep{PC2018, CP2019}, \citet{CC2022} suggested that they may also trace the merging zone between an accretion streamer and a disk. Our simulations support this interpretation, demonstrating that streamer accretion can produce Doppler-flip signatures in moment-1 residual maps. Once infall ceases, the Doppler flip disappears. Nevertheless, velocity residuals with amplitudes of up to $\sim\pm100\ms$ are still observed after the infall has stopped, which allows us to predict the infall history of the observed disk.

\subsection{Position-Velocity Diagram} \label{subsec:pv}

\begin{figure*}[htb!]
    \epsscale{1.0}
    \plotone{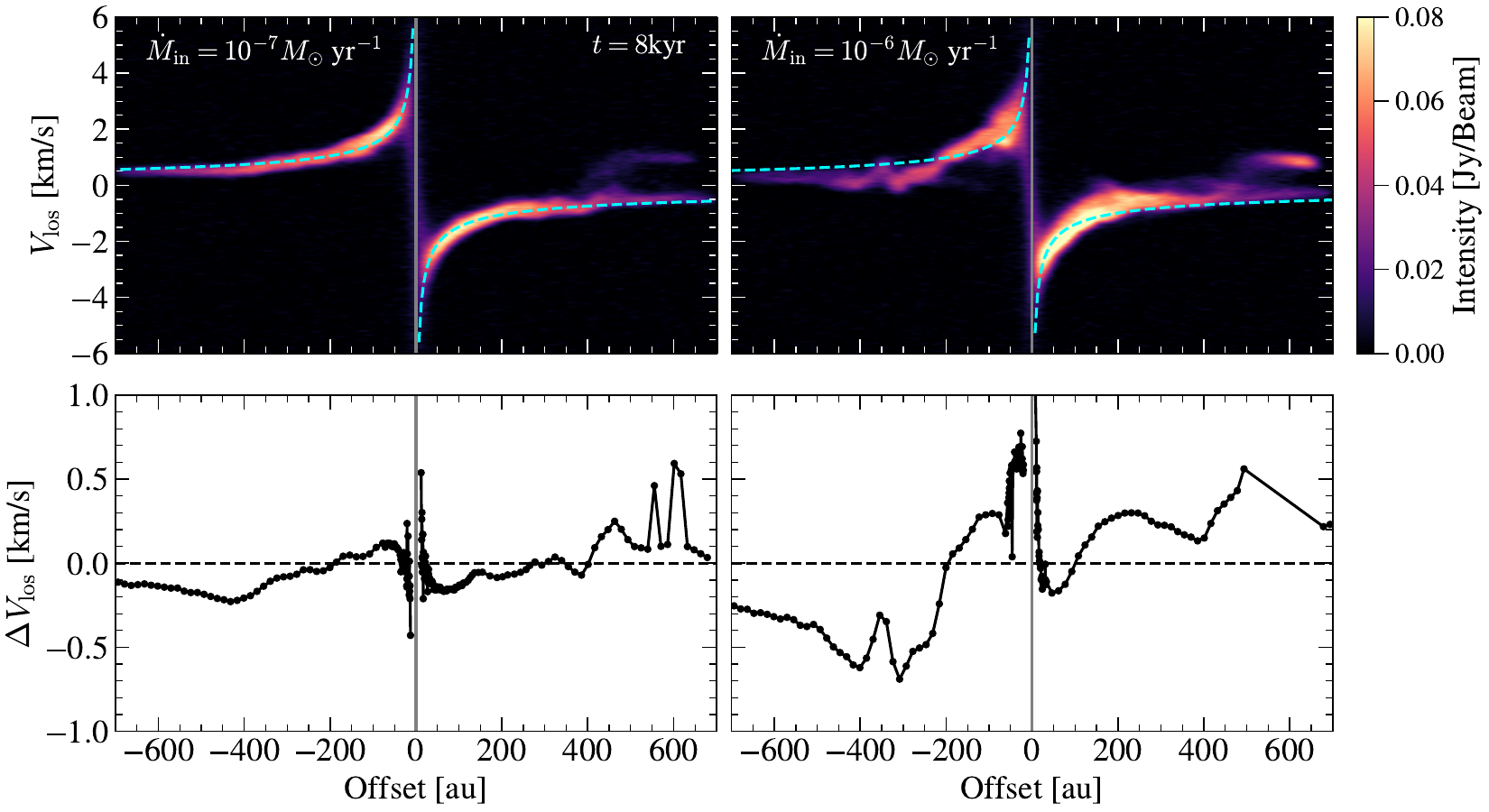}
    \caption{Radial PV diagrams (\textit{top}) and the deviations of the rotation curves measured with \textsc{SLAM} from the best-fitting Keplerian rotation curves (\textit{bottom}) for the synthetic $^{13}$CO observations of the $\dot{M}_{\mathrm{in}}=10^{-7}\mdotyr$ (\textit{left}) and $10^{-6}\mdotyr$ (\textit{right}) models.  Both models exhibit a characteristic PV wiggle as well as a non-disk kinematic feature near $r\sim 600\au$. As the infall rate increases, the PV diagram becomes increasingly asymmetric owing to the combined effects of disk eccentricity and disk warping.}
    \label{fig:pv_major}
\end{figure*}

\begin{figure*}[htb!]
    \epsscale{1.0}
    \plotone{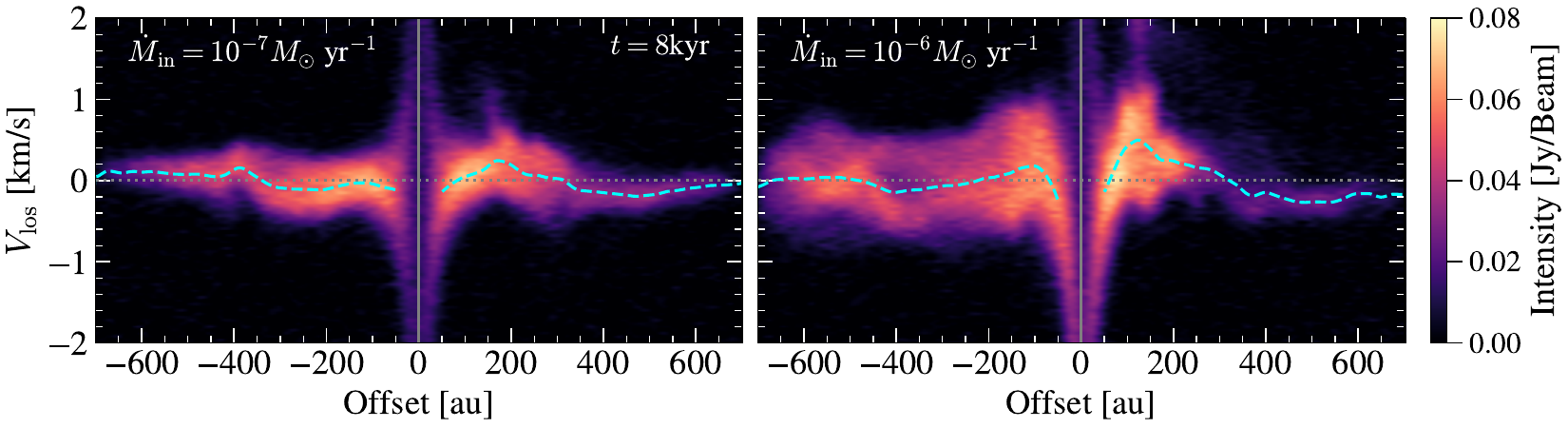}
    \caption{Radial PV diagrams of the synthetic $^{13}$CO observations along the disk minor axis for the $\dot{M}_{\mathrm{in}}=10^{-7}\mdotyr$ (\textit{left}) and $10^{-^6}\mdotyr$ (\textit{right}) models. The centroids of the velocity profiles are shown as cyan dashed lines. The resulting PV wiggles trace velocity perturbations induced by streamer-driven spiral arms.}
    \label{fig:pv_minor}
\end{figure*}

We also investigate the position--velocity (PV) diagrams during the active infall phase. Figures~\ref{fig:pv_major} and \ref{fig:pv_minor} present the PV diagrams at $t=8\kyr$ for the $\dot{M}_{\mathrm{in}}=10^{-7}\mdotyr$ and $10^{-6}\mdotyr$ models, extracted along the disk major and minor axes, respectively. Along the major axis, both models exhibit the characteristic signature of Keplerian rotation. We use \textsc{SLAM}\footnote{https://github.com/jinshisai/SLAM} \citep{AS2024} to extract the rotation curves from the PV diagrams and fit them with a Keplerian rotation profile to estimate the central stellar mass. 

Adopting the disk inclination derived from \textsc{discminer} (Section \ref{subsec:channel}), we obtain $M_*\sim 2.17\msun$ for $\dot{M}_{\mathrm{in}}=10^{-7}\mdotyr$ model, slightly below the true stellar mass of $2.4\msun$. Nevertheless, the best-fitting Keplerian model fails to reproduce the kinematics of the outer disk ($R > 400\au$), where velocity deviations of $\sim 0.2\kms$ are present. This corresponds to a discrepancy of more than $40\%$ in the projected line-of-sight velocity $V_{\mathrm{los}}$. Of the two warp parameters discussed in Section \ref{subsec:angmom}, namely the tilt angle $\beta$ and the twist angle $\gamma$, the latter is the more likely origin of these deviations because its radial variation closely resembles that of the velocity residuals.

The streamer-induced signatures become substantially stronger in the $\dot{M}_{\mathrm{in}}=10^{-6}\mdotyr$ model, shown in the right column of Figure~\ref{fig:pv_major}. In particular, the combination of strong disk eccentricity and disk warping produces a markedly asymmetric PV diagram. As a result, the stellar-mass estimate becomes significantly biased, yielding $M_*\sim 1.7\msun$, substantially lower than the true value.

We also identify a distinct non-disk kinematic component near $r\approx-600$ to $-400\au$, which originates from the primary spiral arm generated by streamer accretion. Similar non-disk features have been reported in the PV diagram of AB Aur, where infall is believed to be ongoing \citep{SD2025}. Our results therefore support the interpretation that such features can serve as observational signatures of active streamer accretion. 

In the PV diagrams extracted along the minor axis, we identify a characteristic ``PV wiggle'' with an amplitude of $\sim 0.2\kms$. Such features have previously been interpreted as a signature of gravitational instability \citep{SD2024}. Our simulations, however, demonstrated that streamer-induced spiral arms can generate similar kinematic perturbations. Therefore, the presence of a PV wiggle alone does not uniquely identify the physical origin of the spiral structure. Caution is needed when distinguishing between gravitational instability, planet--disk interactions, and streamer accretion as the driving mechanism of spiral arms in observed disks.

\section{Discussion} \label{sec:discuss}

\subsection{Position of Streamers} \label{subsec:vpos}
\begin{figure*}[htb!]
    \epsscale{1.0}
    \plotone{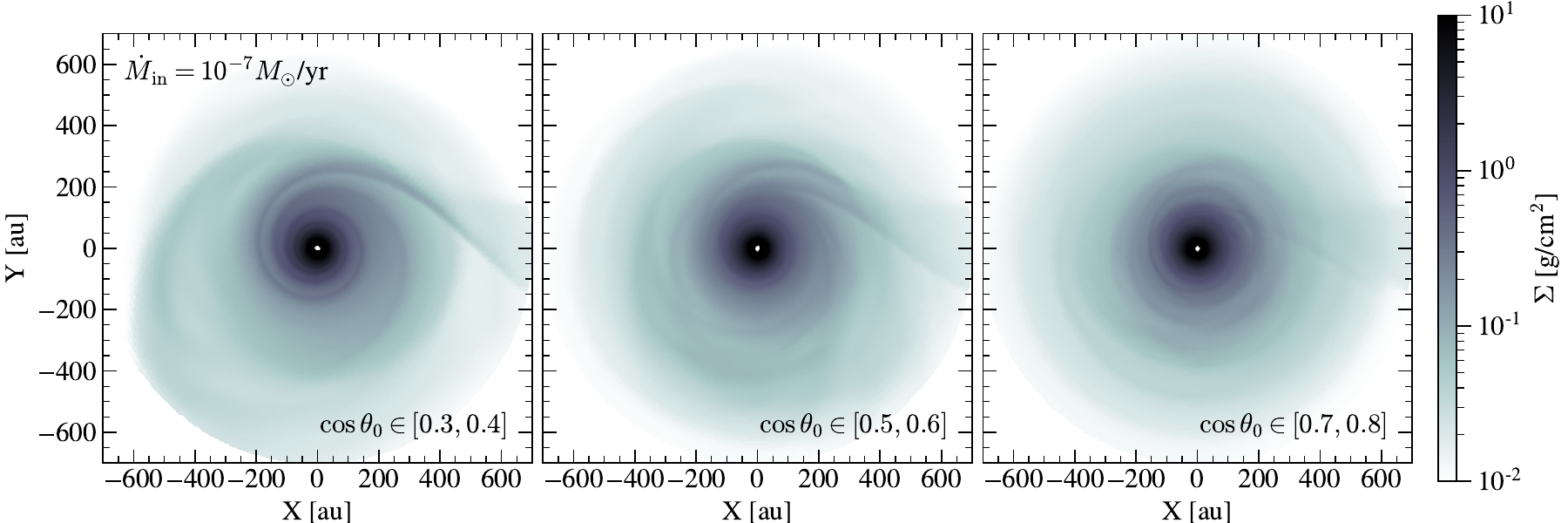}
    \caption{Snapshots of the surface density at the end of the infall phase ($t=8\kyr$) for the $\dot{M}_{\mathrm{in}}=10^{-7}\mdotyr$ models with $\cos\theta_0\in[0.3,0.4]$ (\textit{left}), $[0.5,0.6]$ (\textit{middle}), and $[0.7,0.8]$ (\textit{right}). As the streamer approaches the disk midplane (smaller $\cos\theta_0$), the resulting spiral arm becomes more extended and prominent in the outer disk.}

    \label{fig:position}
\end{figure*}

We further investigate how the disk response depends on the streamer's injection angle relative to the disk. To this end, we run additional simulations in which we vary the range of $\cos\theta_0$ defining the infall region, thereby changing the streamer's vertical location. Figure~\ref{fig:position} compares the surface density distributions at $t=8\kyr$ for the $\dot{M}_{\mathrm{in}}=10^{-7}\mdotyr$ models with different ranges of $\cos\theta_0$.

\begin{figure}[htb!]
    \epsscale{1.15}
    \plotone{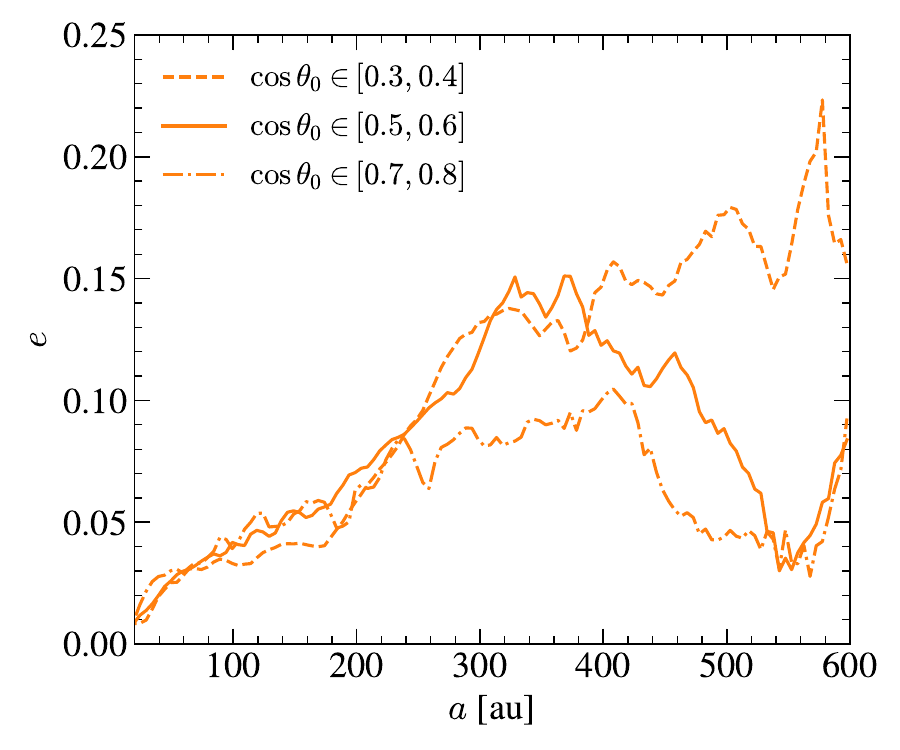}
    \caption{Eccentricity measured in each semi-major axis bin for the $\dot{M}_{\mathrm{in}}=10^{-7}\mdotyr$ model at $t=8$ kyr, shown for different streamer injection angles: $\cos\theta_0\in[0.3,0.4]$ (\textit{dashed}), $[0.5,0.6]$ (\textit{solid}), and $[0.7,0.8]$ (\textit{dot-dashed}). As the streamer approaches the disk midplane (smaller $\cos\theta_0$), its influence extends to larger radii, resulting in stronger eccentricity excitation in the outer disk.}
    \label{fig:position_eccen}
\end{figure}

In general, the impact of the streamer shifts toward larger radii as its injection point approaches the disk midplane (see \cref{fig:diskmodel}). Figure~\ref{fig:position_eccen} plots the radial distribution of disk eccentricity at $t=8\kyr$ for different ranges of $\cos\theta_0$. The streamer with the smallest value of $\cos\theta_0$ excites significant eccentricity in the outer disk ($a\gtrsim400\au$), whereas streamers with larger values of $\cos\theta_0$ primarily excite eccentricity at smaller radii ($a\lesssim300\au$).
In addition, the overall level of eccentricity excitation decreases as the streamer becomes more perpendicular to the disk midplane. As a result, the associated disk substructures become progressively weaker, consistent with the surface density distributions shown in Figure~\ref{fig:position}. Despite these differences in the outer disk, the eccentricity excited in the inner disk remains broadly similar among the models and therefore produces only minor differences in the accretion rate.

\begin{figure}[htb!]
    \epsscale{1.15}
    \plotone{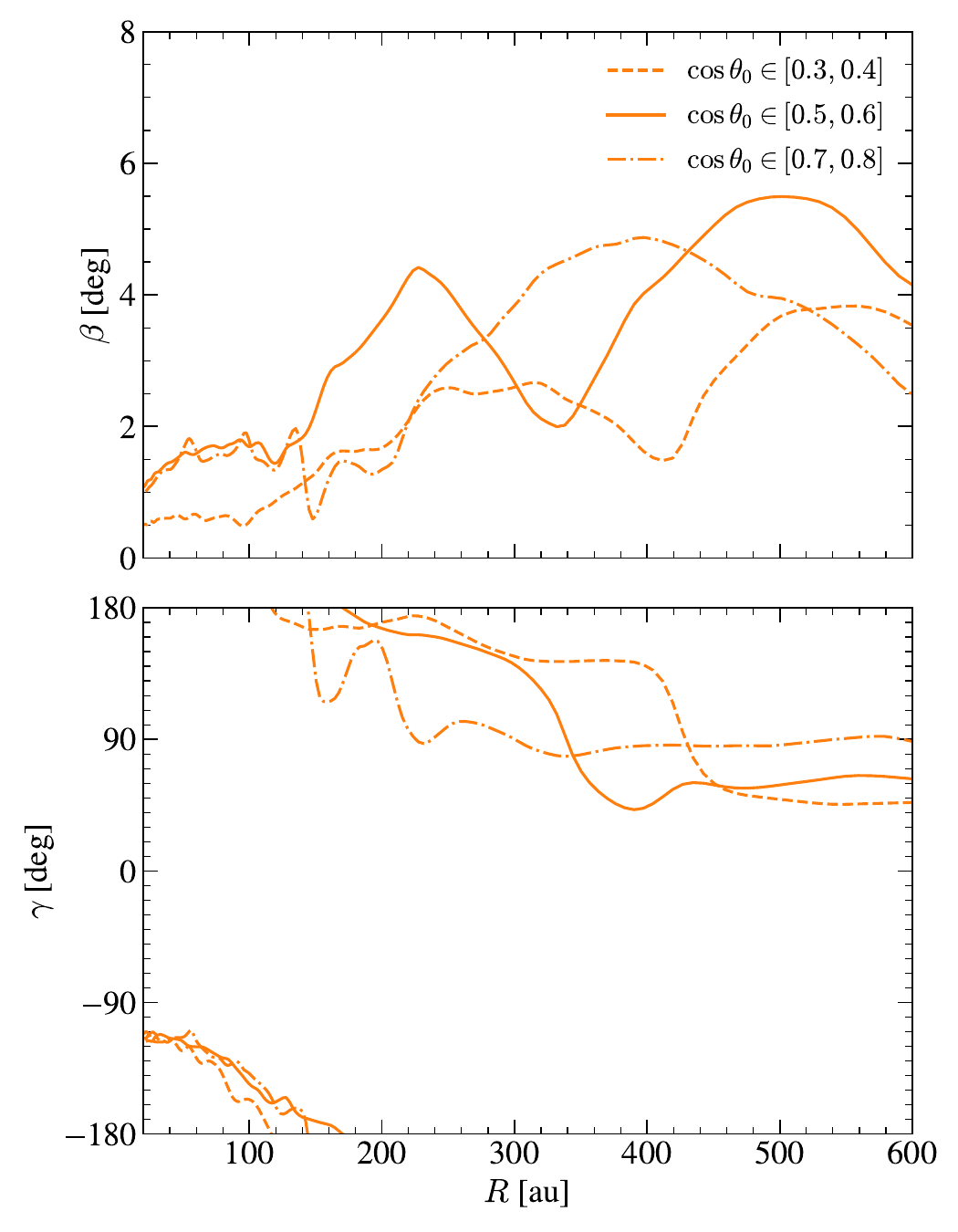}
    \caption{Angular momentum parameters measured for $\dot{M}_{\mathrm{in}}=10^{-7}\mdotyr$ at $t=8$kyr for streamers with the vertical position of $\cos\theta_0\in[0.3,0.4]$ (\textit{dashed}), $[0.5,0.6]$ (\textit{solid}), and $[0.7,0.8]$ (\textit{dotted-dashed}). The streamer-induced tilt exhibits a non-monotonic dependence on the injection angle because it is governed by both the direction of the infalling streamer and the total angular momentum it carries.}
    \label{fig:vpos_angmom}
\end{figure}

Conversely, the effect of streamer accretion on the orientation of the disk angular momentum vector becomes weaker as the streamer approaches the disk midplane, since the angular momentum of the infalling material becomes more closely aligned with that of the disk. Figure~\ref{fig:vpos_angmom} plots the radial profiles of the tilt ($\beta$) and twist ($\gamma$) measured at the end of the infall phase for different streamer injection angles. We find that the smallest tilt occurs for the $\cos\theta_0\in[0.3,0.4]$ model.

Interestingly, the streamer-induced tilt does not increase monotonically with the streamer's vertical elevation. This is because the total angular momentum injection rate decreases as the streamer originates farther from the disk midplane. As illustrated by the streamline geometry in the left panel of \cref{fig:diskmodel}, streamers launched at higher latitudes intersect the disk at smaller radii and therefore carry less angular momentum. Consequently, the magnitude of the induced disk tilt is governed not only by the misalignment between the angular momentum vectors of the disk and the infalling material, but also by the total angular momentum delivered by the streamer. 

\subsection{Multiple Streamers} \label{subsec:multi}

Although this study primarily focuses on disks accreting through a single streamer, observations frequently reveal systems fed by multiple streamers \citep{PA2023, SJ2025}. It is therefore important to assess whether the disk's dynamical response differs between single-streamer and multi-streamer accretion.

To this end, we run additional simulations with two accretion streamers. The first streamer matches that adopted in our fiducial model ($\cos\theta_0\in[0.5,0.6]$, $\Delta\phi=0.4$). A second streamer of equal angular extent is then introduced at an azimuthal separation of $180^\circ$. For its vertical location, we consider two configurations: an Up--Up model, in which both streamers are located above the disk midplane ($\cos\theta_0\in[0.5,0.6]$), and an Up--Down model, in which the second streamer is located below the disk midplane ($\cos\theta_0\in[-0.6,-0.5]$). In all cases, the two streamers are assigned equal mass infall rates, such that the total infall rate, $\dot{M}_{\mathrm{in,total}}$, is identical to that of the corresponding single-streamer simulation. Although many additional configurations are possible, differing in the streamers' relative orientation, location, and strength, we restrict our analysis to these two representative cases. We defer a more comprehensive exploration of the multi-streamer parameter space to future work.

\begin{figure}[htb!]
    \epsscale{1.0}
    \plotone{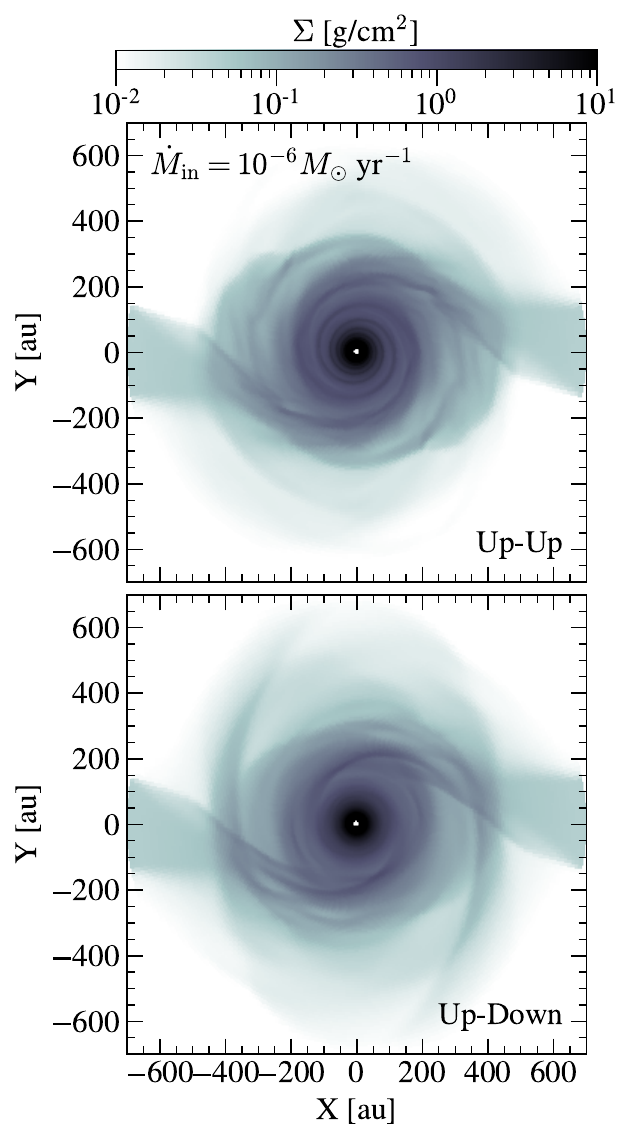}
    \caption{Snapshots of the surface density at $t=8\kyr$ for models with a total mass infall rate of $\dot{M}_{\mathrm{in,total}}=10^{-6}\mdotyr$. The two streamers are either located on the same side of the disk (Up--Up; \textit{top}) or on opposite sides of the disk (Up--Down; \textit{bottom}).}
    \label{fig:multi_surf}
\end{figure}

\begin{figure}[htb!]
    \epsscale{1.15}
    \plotone{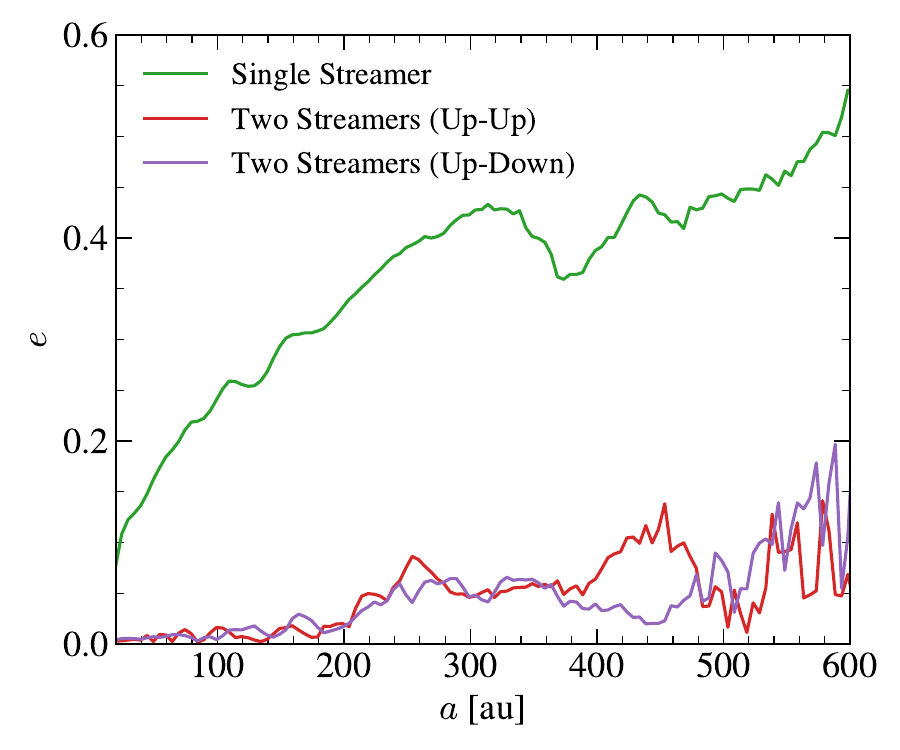}
    \caption{Disk eccentricity at $t=8\kyr$ for models with a total mass infall rate of $\dot{M}_{\mathrm{in,total}}=10^{-6}\mdotyr$, comparing the single-streamer (\textit{green}), Up--Up (\textit{red}), and Up--Down (\textit{purple}) configurations. In general, the multi-streamer models excite substantially less eccentricity than the corresponding single-streamer model.}
    \label{fig:multi_eccen}
\end{figure}

Figure~\ref{fig:multi_surf} shows snapshots of the disk surface density at $t=8\kyr$ for models with a total mass infall rate of $\dot{M}_{\mathrm{in,total}}=10^{-6}\mdotyr$. The most striking difference between the multi-streamer models and the corresponding single-streamer model (Figure~\ref{fig:spiralsnapshot}) is the much weaker eccentricity excitation in the former. Owing to the greater symmetry of the infall geometry, the perturbations induced by individual streamers partially cancel one another, thereby reducing the net eccentricity growth. We quantify this effect in Figure~\ref{fig:multi_eccen}, which compares the radial eccentricity profiles of the multi-streamer and single-streamer models. The eccentricity excited in the multi-streamer simulations is substantially lower than that in the single-streamer case. However, this does not imply that streamer accretion is generally less dynamically influential when multiple streamers are present. Rather, it demonstrates that the disk response depends sensitively on the number, geometry, and relative orientations of the streamers, which together determine the net angular momentum and energy injected into the disk.

Next, we examine the structural differences between the two multi-streamer configurations. In the Up--Up model, where both streamers enter from the same side of the disk, the infalling material excites two prominent primary spiral arms in the outer disk as well as weak $m=2$ spiral density waves in the inner disk. By contrast, the Up--Down model exhibits only the two primary spiral arms, without any discernible inner $m=2$ spiral density waves. 
In the Up--Down configuration, the interaction between the two opposing streamers generates stronger turbulence, disrupting the coherent inner spiral pattern seen in the Up--Up case. Instead, elongated tail-like structures develop on opposite sides of the disk. These features do not evolve into the crescent-shaped overdensities found in the single-streamer model because the excitation of disk eccentricity is substantially weaker.

\begin{figure*}[htb!]
    \epsscale{1.0}
    \plotone{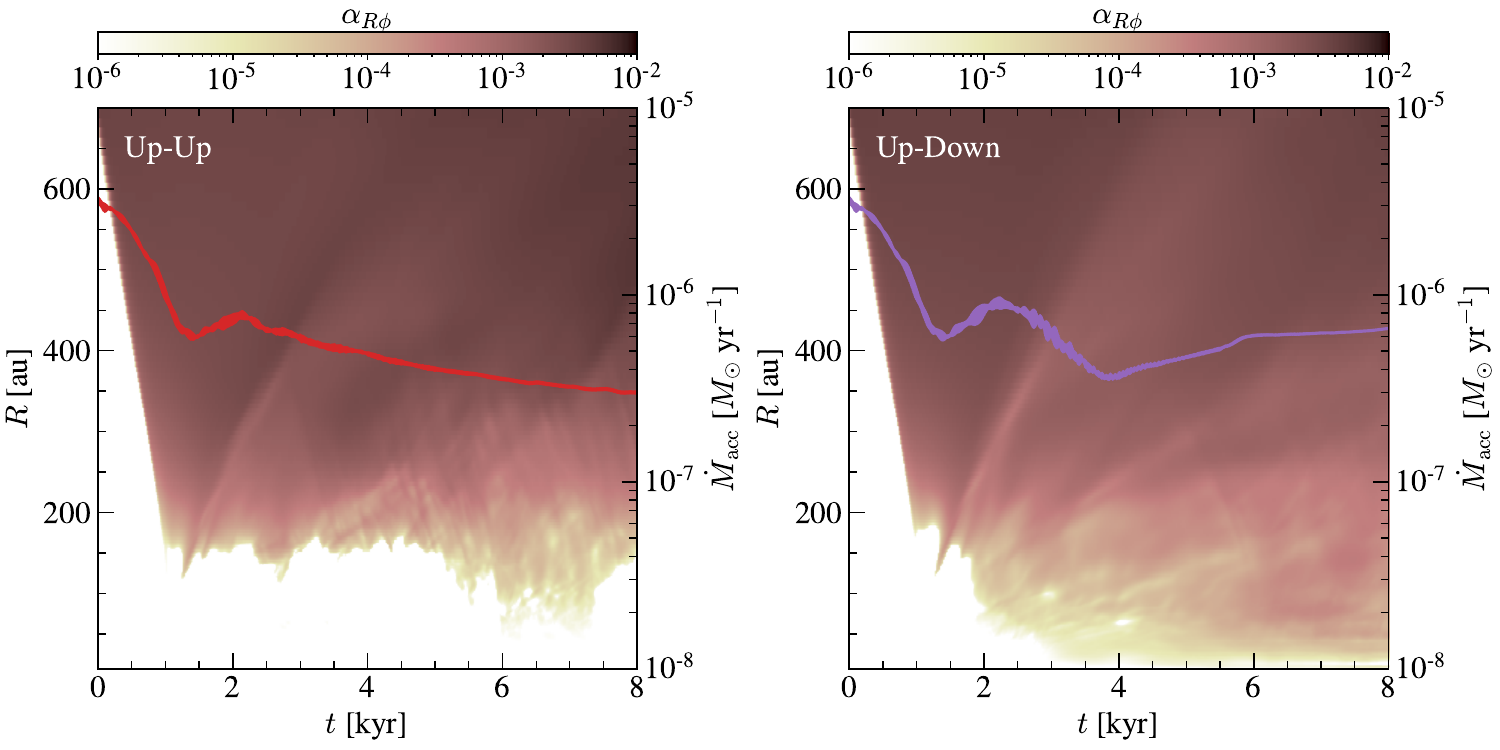}
  \caption{Accretion rates for the multi-streamer simulations with a total mass infall rate of $\dot{M}_{\mathrm{in,total}}=10^{-6}\mdotyr$ in the Up--Up (\textit{left}) and Up--Down (\textit{right}) configurations. The pressure-weighted Reynolds stress, $\alpha_{R\phi}$, is shown in the background. Unlike the single-streamer model (Figure~\ref{fig:accretionrate}), neither configuration produces a strong accretion outburst. However, the enhanced turbulence generated in the Up--Down model leads to a final accretion rate approximately twice that of the Up--Up model.}
    \label{fig:multi_accrate}
\end{figure*}

The relative orientation of the second streamer also affects the accretion history of the central star. \cref{fig:multi_accrate} shows the temporal evolution of the accretion rate during the active infall phase. Because eccentricity growth remains weak in both multi-streamer models, neither exhibits the strong accretion outbursts seen in the single-streamer case (Figure~\ref{fig:accretionrate}). Nevertheless, the final accretion rate in the Up--Down model is approximately twice that in the Up--Up model.

To investigate the origin of this difference, we compute the pressure-weighted Reynolds stress,
\begin{equation}
    \alpha_{R\phi}(R)\equiv \frac{\int \rho v_R \delta v_{\phi} \, d\phi d\theta}{\int P \, d\phi d\theta}\,,
\end{equation}
where $\delta v_{\phi}\equiv v_{\phi}-\langle v_{\phi}\rangle$. The resulting values of $\alpha_{R\phi}$ are shown alongside the accretion rates in \cref{fig:multi_accrate}. We find that the Up--Down configuration consistently produces larger Reynolds stresses than the Up--Up configuration, indicating higher turbulence in the disk. The enhanced turbulence in the Up--Down model likely reflects the fact that the two streamers perturb opposite sides of the disk simultaneously, producing stronger velocity fluctuations throughout the disk volume. In contrast, the Up--Up configuration primarily disturbs the upper disk layers, resulting in weaker turbulent transport and a correspondingly lower accretion rate.

\begin{figure}[htb!]
    \epsscale{1.0}
    \plotone{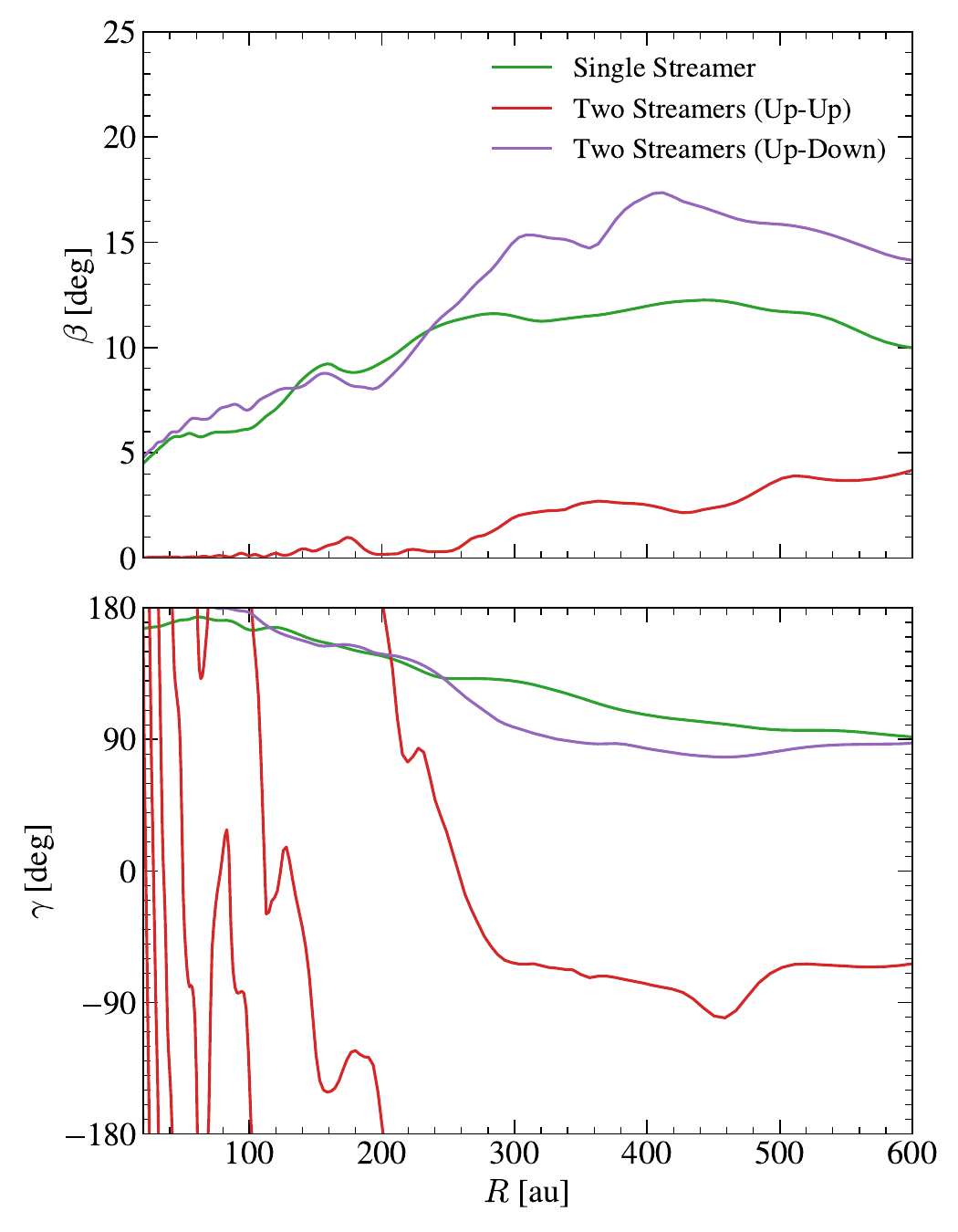}
    \caption{Angular momentum parameters measured at $t=8\kyr$ for models with a total mass infall rate of $\dot{M}_{\mathrm{in,total}}=10^{-6}\mdotyr$, comparing the single-streamer (\textit{green}), Up--Up (\textit{red}), and Up--Down (\textit{purple}) configurations. The resulting tilt and twist profiles depend sensitively on the streamers' relative orientation, showing that the disk's angular-momentum response is strongly influenced by infall geometry.}
    \label{fig:multi_angmom}
\end{figure}

Finally, we examine how multiple streamers affect the disk's angular momentum structure. \cref{fig:multi_angmom} shows the radial profiles of the disk tilt ($\beta$) and twist ($\gamma$) measured at the end of the infall phase ($t=8\kyr$) for the multi-streamer models, together with the corresponding single-streamer result. Relative to the single-streamer model, the Up--Down configuration produces a larger tilt in the outer disk, while leaving the overall twist profile largely unchanged. In contrast, the Up--Up configuration induces a substantially smaller tilt because the angular momentum injected by the two streamers is more symmetrically distributed with respect to the disk midplane. The twist profile in this model, however, differs noticeably from the single-streamer case and shows strong fluctuations in the inner disk. These fluctuations are associated with the prominent two-armed spiral pattern shown in \cref{fig:multi_surf}, which introduces local variations in the angular momentum distribution.

\subsection{Retrograde Streamer}\label{subsec:retro}

Although the origin of accretion streamers remains uncertain \citep{PA2023}, their formation is likely driven by external processes, implying that the infalling material may possess either prograde or retrograde angular momentum relative to the disk. Recent numerical studies have demonstrated the existence of retrograde streamers and highlighted their distinct effects on disk evolution \citep{KP2024, GD2024}. Motivated by these findings, we investigate how the impact of a retrograde streamer differs from that of the prograde streamer considered throughout the rest of this study. To construct the retrograde-streamer model, we reverse the sign of the azimuthal velocity, $v_{\phi}$ (\cref{eq:ucmvaz}), while keeping all other streamer properties, including its spatial extent and injection geometry, identical to those of the fiducial model.

\begin{figure*}[htb!]
    \epsscale{1.15}
    \plotone{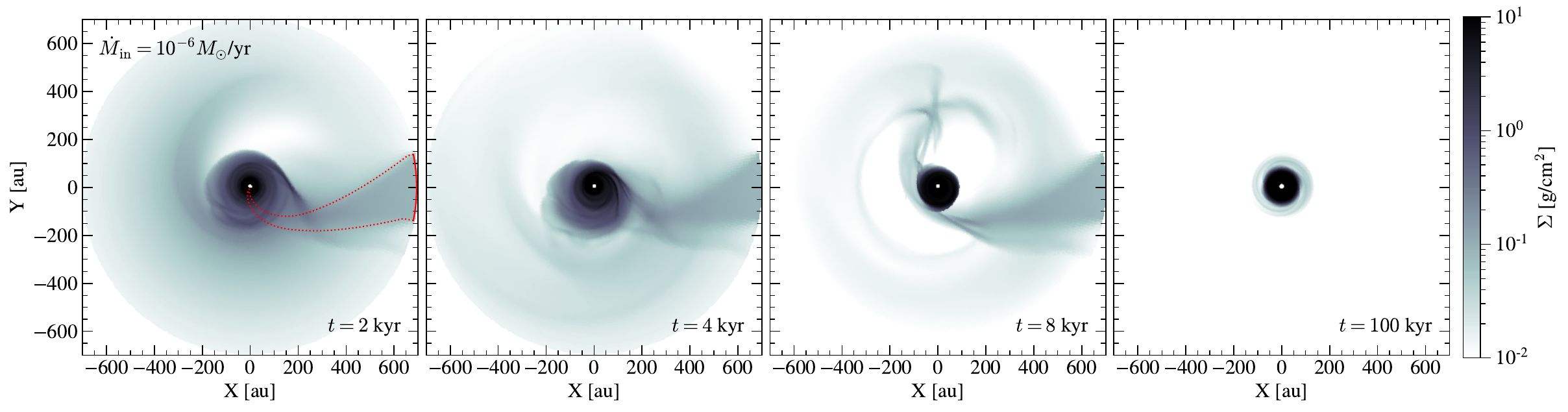}
    \caption{Snapshots of the surface density at $t=2$, $4$, $8$, and $100\kyr$ (from left to right) for the retrograde-streamer model with $\dot{M}_{\mathrm{in}}=10^{-6}\mdotyr$. The infall region and streamer trajectories are overplotted in the same manner as in Figure~\ref{fig:spiralsnapshot}. Retrograde streamer accretion efficiently removes angular momentum from the disk, producing a compact disk with a characteristic radius of $R\lesssim100\au$.}
    \label{fig:retro_snapshot}
\end{figure*}

\cref{fig:retro_snapshot} plots snapshots of the disk surface density during the infall phase ($t=2$--$8\kyr$) and after the cessation of infall ($t=100\kyr$) for the retrograde-streamer model with $\dot{M}_{\mathrm{in}}=10^{-6}\mdotyr$. The most striking consequence of retrograde streamer accretion is the dramatic reduction in disk size. As indicated by \cref{eq:angmominjrate}, the total angular momentum delivered by the streamer over $8\kyr$ corresponds to $\sim55\%$ of the initial angular momentum of the disk. Because this angular momentum is anti-aligned with the disk rotation, it effectively cancels a substantial fraction of the disk's original angular momentum, causing the disk to contract to a characteristic radius of $\lesssim100\au$. After infall ceases, the disk separates into two distinct components: a compact inner disk ($R\lesssim100\au$) and a diffuse outer component ($R\gtrsim100\au$). The latter consists primarily of material originating from the infalling streamer, which remains only weakly bound to the compact inner disk.

\begin{figure}[htb!]
    \epsscale{1.15}
    \plotone{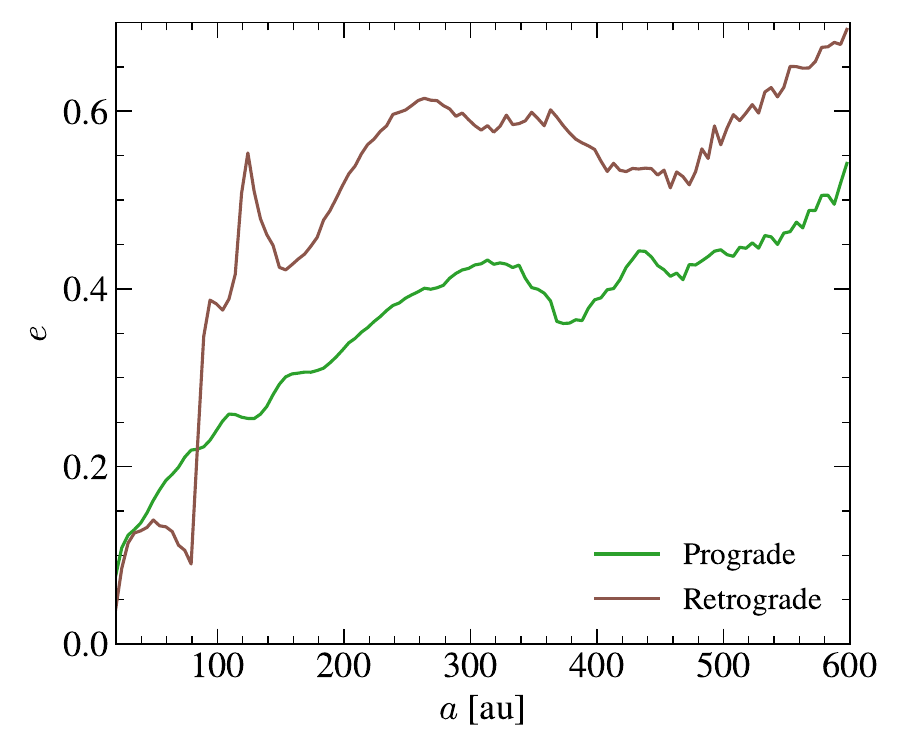}
    \caption{Disk eccentricity at $t=8\kyr$ for models with an infall rate of $\dot{M}_{\mathrm{in}}=10^{-6}\mdotyr$, comparing the prograde-streamer (\textit{green}) and retrograde-streamer (\textit{brown}) cases. In the retrograde-streamer model, the eccentricity profile exhibits a pronounced transition near $R\sim100,\au$, separating a weakly eccentric inner disk from a highly eccentric outer component.}
    \label{fig:retro_eccen}
\end{figure}

We examine the eccentricity excited by the retrograde streamer. \cref{fig:retro_eccen} compares the disk eccentricity at the end of the infall phase ($t=8\kyr$) with that of the fiducial prograde-streamer model. In the retrograde case, the eccentricity profile exhibits a pronounced transition near $R\sim100\au$, separating the disk into dynamically distinct inner and outer regions.
Beyond $R\sim100\au$, the disk surface density is strongly depleted owing to the loss of angular momentum induced by the retrograde infall. As a result, the orbital structure in this region is dominated by the streamer material itself, which follows highly eccentric, nearly parabolic trajectories. In contrast, the eccentricity of the outer component does not efficiently propagate inward. The sharp pressure enhancement associated with the truncated disk edge acts as a barrier to the propagation of eccentricity-carrying waves, thereby suppressing angular momentum and eccentricity exchange between the inner and outer regions. Consequently, the compact inner disk remains only weakly eccentric, in marked contrast to the prograde-streamer model, where eccentricity grows throughout a much larger fraction of the disk.

\begin{figure*}[htb!]
    \epsscale{1.00}
    \plotone{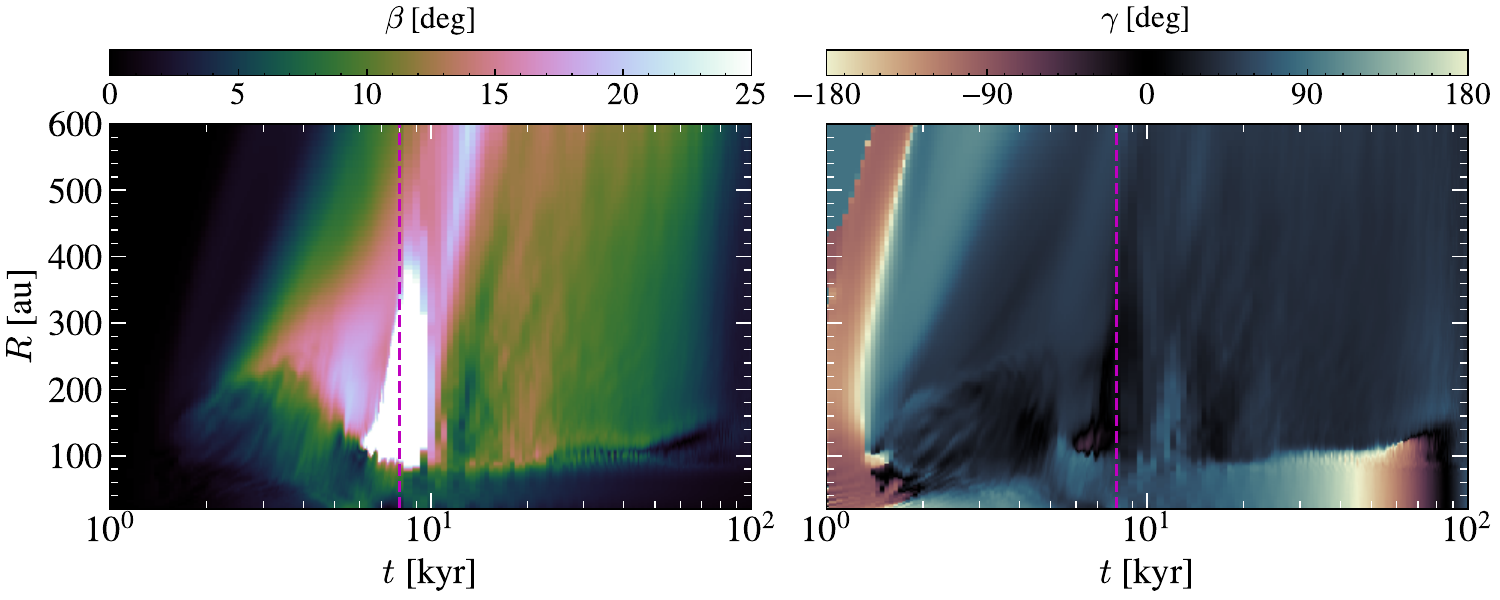}
    \caption{Temporal evolution of the angular momentum parameters $\beta$ (\textit{left}) and $\gamma$ (\textit{right}) for the retrograde-streamer model with $\dot{M}_{\mathrm{in}}=10^{-6}\mdotyr$. A sharp increase in $\beta$ at $100\au\lesssim R\lesssim400\au$ between $t\sim6$ and $10\kyr$, together with the distinct angular momentum evolution of the inner ($R\lesssim100,\au$) and outer ($R\gtrsim100,\au$) regions, suggests the onset of disk breaking.}
    \label{fig:retro_angmom}
\end{figure*}

\begin{figure*}[htb!]
    \epsscale{1.00}
    \plotone{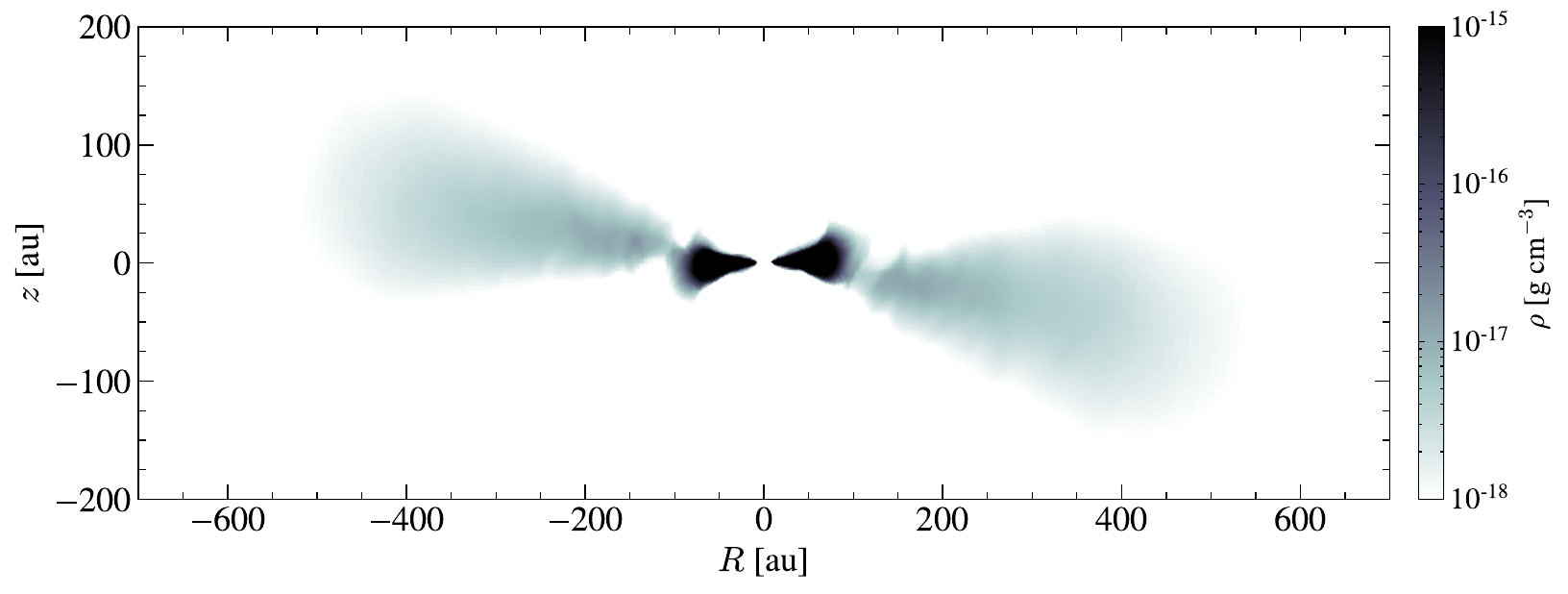}
    \caption{Two-dimensional slice of the gas density in the meridional ($R$--$z$) plane passing through $\phi=0^\circ$ and $180^\circ$. A clear discontinuity in the disk inclination is visible near $R\sim100\,\mathrm{au}$, indicating the onset of disk breaking.}
    \label{fig:retro_dens}
\end{figure*}

We find that the retrograde streamer most strongly affects disk angular momentum evolution, ultimately leading to disk breaking. \cref{fig:retro_angmom} shows the temporal evolution of $\beta$ and $\gamma$ for the retrograde-streamer model with $\dot{M}_{\mathrm{in}}=10^{-6}\mdotyr$. A pronounced increase in $\beta$ develops over $100\au \lesssim R \lesssim 400\au$ between $t\sim6$ and $10\kyr$. This feature is primarily caused by the severe depletion of disk material in this region, which makes the local angular momentum vector poorly defined, rather than by a genuine increase in disk inclination.
In addition, the markedly different evolution of $\beta$ and $\gamma$ in the inner ($R\lesssim100\au$) and outer ($R\gtrsim100\au$) regions indicates that these two components become dynamically decoupled. Such behavior is consistent with streamer-induced disk breaking.

\cref{fig:retro_dens} provides further evidence for this interpretation by showing the density structure of the disk after the retrograde infall has ceased. A clear discontinuity in the disk inclination is visible near $R\sim100\,\au$. The diffuse outer component is warped by $\beta\sim10^\circ$ relative to the dense inner component and exhibits coherent vertical bending \citep[see e.g.,][]{NP1999}, while the inner disk remains nearly planar. This sharp change in the disk orientation is characteristic of the onset of disk breaking.

\begin{figure}[htb!]
    \epsscale{1.15}
    \plotone{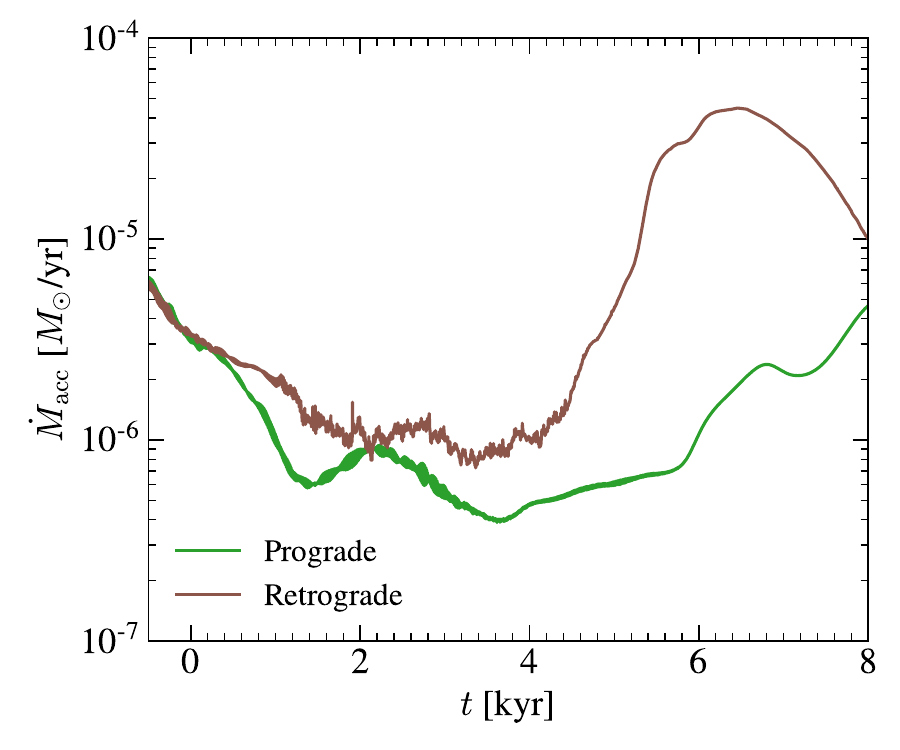}
    \caption{Accretion rates during the infall phase for the prograde-streamer (\textit{green}) and retrograde-streamer (\textit{brown}) models, both with $\dot{M}_{\mathrm{in}}=10^{-6}\mdotyr$. Compared to the prograde case, retrograde streamer accretion produces a more rapid and stronger enhancement of the stellar accretion rate.}
    \label{fig:retro_accrate}
\end{figure}

Finally, we compare the accretion histories of the prograde- and retrograde-streamer models during the infall phase, as shown in \cref{fig:retro_accrate}. We find that retrograde streamer accretion produces a systematically higher stellar accretion rate than the corresponding prograde case. In addition, the accretion rate rises more rapidly in the retrograde model, reflecting the efficient removal of disk angular momentum by the counter-rotating infalling material. Although the resulting rise time remains longer than that typically observed in YSO accretion outbursts, these results suggest that streamer accretion may nevertheless contribute to rapid accretion variability. In particular, if the merging zone between the streamer and the disk were located much closer to the central star, the resulting angular momentum loss and mass transport could occur on much shorter timescales, potentially providing a pathway to explain observed accretion outburst events.

\subsection{Implications} \label{subsec:implication}

A primary consequence of streamer accretion onto a PPD is the global excitation of disk eccentricity. Previous studies of eccentric PPDs have focused primarily on eccentricity driven by stellar companions \citep{LM2023} or embedded massive planets \citep{TO2016}. In these systems, eccentric Lindblad resonances generate eccentricity \citep{L1991}. For example, stellar binaries excite eccentricity in the inner regions of circumbinary disks \citep{MM2008, MM2017}, whereas massive planets primarily drive eccentricity near the outer edge of the gap they carve in the disk \citep{TO2017, TK2022}.
In contrast, streamer accretion excites eccentricity over a much broader radial extent. As a result, caution is required when directly applying results derived for planet- or binary-driven eccentric disks to the streamer-driven case. Nevertheless, if the disk is approximated as a collection of coplanar gas rings on eccentric orbits \citep{OB2014}, many of the dynamical consequences of eccentricity can still be understood qualitatively. This framework provides a useful basis for interpreting how streamer-driven eccentricity modifies the disk's internal evolution and dynamics.

In the context of planet formation, \citet{H1978} originally argued that large eccentricities can inhibit planetesimal growth by increasing collision velocities, making collisions destructive rather than accretional. Nevertheless, the prevalence of exoplanets in systems exhibiting substantial eccentricities \citep{HC2003, DC2011} demonstrates that planet formation can proceed even in dynamically excited environments. One mechanism for overcoming this fragmentation barrier is gas drag, which can apsidally align planetesimal orbits and thereby reduce their relative velocities \citep{MS2000}. In addition, disk self-gravity can suppress the secular excitation of planetesimal eccentricities \citep{SR2015}. Taking these effects into account, \citet{SR2021} showed that planet formation remains viable provided that the initial planetesimals are larger than $\sim10$ km, a size scale that can be readily produced through processes such as the streaming instability.

Planet--disk interactions are also expected to be fundamentally modified in eccentric disks. For example, \citet{SR2025} recently showed that massive-planet migration can stall or even reverse in sufficiently eccentric disks, in contrast to the classical picture of steady inward migration \citep{PL1984, SC1995}. They demonstrated that this behavior is driven by eccentricity excited near the outer edge of the planet-induced gap, which broadens the gap and substantially weakens the negative torque exerted by the outer disk.
Given that streamer accretion excites eccentricity over a much broader radial extent, its impact on planet migration could be even more pronounced. In particular, streamer-driven eccentricity may alter gap structures, modify the balance of gravitational torques acting on embedded planets, and ultimately drive migration pathways that differ substantially from those expected in the standard Type II migration framework.

\subsection{Caveats} \label{subsec:caveat}

In this study, we modeled accretion streamers using a modified UCM framework. This approach provides a convenient way to control the streamer's physical properties and systematically explore its impact on disk evolution. However, it remains a highly idealized representation of the interaction between a PPD and its surrounding environment compared to fully self-consistent environmental simulations \citep{KP2024, HK2026}. In addition, the classical UCM model assumes that the infalling material originates from an envelope of effectively infinite extent and possesses zero initial radial velocity \citep{U1976, CM1981}. These assumptions may not accurately describe realistic star-forming environments. To address this limitation, \citet{PS2020} developed an extended streamer model based on the formalism of \citet{MT2009}, which relaxes several of the assumptions underlying the classical UCM solution. Incorporating such a framework would provide a more realistic description of streamer kinematics and therefore represents a promising direction for future work.

Additionally, our hydrodynamic simulations neglect magnetic fields, which may play an important role in regulating streamer accretion. For example, \citet{UH2022} demonstrated that magnetic fields can substantially alter the morphology and dynamics of accretion channels formed through cloudlet capture. More recently, molecular-line observations of SVS~13A by \citet{CPH2025} revealed a sub-Alfv\'{e}nic streamer whose kinematics appear to be guided by magnetic field lines. Because our streamer model is based on the UCM formalism, which assumes purely gravitational, hydrodynamic free fall toward the central star, some of our predictions regarding streamer-induced disk substructures, kinematics, and angular momentum transport may require revision once magnetic fields are included.

Another important limitation is the absence of localized shock heating at the streamer--disk interface. In our simulations, shock heating at the merging zone is largely suppressed because the cooling timescale given by \cref{eq:tcool} is short in the outer disk. Consequently, gas that is heated by the impact rapidly relaxes back toward the prescribed equilibrium temperature, preventing the formation of strong localized temperature enhancements associated with accretion shocks. This limitation is particularly relevant because molecular shock tracers such as SO, SO$_2$, and H$_2$CO are routinely used to identify and characterize active accretion streamers \citep{GP2022, LMK2023, SD2025, SJ2025}. A realistic treatment of shock heating is therefore essential for direct comparisons between theoretical models and observations. Future radiation-hydrodynamic simulations \citep{KB2009, BC2013, YK2022} should provide a more self-consistent description of the thermal structure of the streamer--disk interaction region and help overcome this limitation.

We also note that our simulation results may be affected by numerical diffusion, as discussed in Appendix \ref{app:amd}. Because this study explores the long-term effects of accretion streamers on PPDs over a range of streamer properties, computational constraints limit the numerical resolution we can use. Previous studies of eccentric disks show that numerical dissipation from eccentric streamlines oblique to the grid can artificially damp disk eccentricity \citep{BO2016, LD2023}. Similarly, \citet{KD2024} showed that insufficient vertical resolution can spuriously decrease disk inclination, although our resolution is a factor of a few higher than that used in \citet{KD2024}. These numerical effects may therefore cause our simulations to underestimate the persistence of streamer-induced disk structures. Future studies should revisit the long-term evolution and longevity of these structures using higher numerical resolution.

\section{Summary} \label{sec:summary}

In this paper, we investigated the dynamical impact of accretion streamers on PPDs and their central protostars using three-dimensional hydrodynamic simulations based on a modified UCM streamer model, together with synthetic ALMA molecular-line observations. Our main findings are summarized as follows.

\begin{enumerate}

\item Accretion streamers efficiently excite global disk eccentricity, with eccentricities reaching values as high as $e\sim0.4$ for the highest infall rates considered. The resulting eccentric disks develop prominent spiral arms and crescent-like overdensities. The spatial distributions of these structures show remarkable agreement with the surface density modulation predicted by the analytic eccentric-disk framework of \citet{OL2019}, supporting their origin in orbital crowding associated with disk eccentricity. The eccentric disks also exhibit enhanced stellar accretion, with the stellar accretion rate increasing by a factor of $\sim30$ relative to the initial state for $\dot{M}_{\mathrm{in}}=10^{-6}\mdotyr$.

\item Streamer accretion modifies the angular momentum structure of the disk. Prograde streamers induce disk warping and generate a modest misalignment between the inner and outer disk. The magnitude of the induced tilt and twist depends sensitively on the geometry of the infalling material.

\item Synthetic $^{13}$CO observations reveal a variety of observable streamer-induced signatures, including residuals in velocity channel maps, Doppler flips in moment-1 maps, and wiggles in position--velocity diagrams. These signatures can persist for up to $\sim100\kyr$ after infall has ceased, suggesting that some kinematic disturbances observed in disks without currently detected streamers may be relics of past infall events.

\item The impact of a streamer depends strongly on its injection geometry. Streamers entering closer to the disk midplane excite stronger eccentricity in the outer disk, whereas high-latitude streamers produce larger disk tilts. Thus, the radial extent of streamer-induced perturbations and the resulting angular momentum evolution are closely tied to the streamer's trajectory.

\item When multiple streamers are present, eccentricity excitation is substantially suppressed relative to the single-streamer case. Nevertheless, the disk response remains highly sensitive to the relative orientations of the streamers. In particular, streamers impacting opposite disk hemispheres generate stronger turbulence and higher stellar accretion rates than streamers entering from the same hemisphere.

\item Retrograde streamer accretion produces a qualitatively different mode of disk evolution from prograde infall. By removing angular momentum from the disk, retrograde streamers generate compact disks and can trigger disk breaking, leading to a dynamical decoupling between the inner and outer disk. They also drive a stronger and more rapid increase in the stellar accretion rate than their prograde counterparts.

\end{enumerate}

The diversity of outcomes found in our parameter survey indicates that the long-term evolution of a PPD can depend sensitively on the properties of ongoing environmental accretion. Our results demonstrate that accretion streamers can profoundly reshape the structure, kinematics, and angular momentum evolution of PPDs. In particular, streamer-driven eccentricity provides a natural mechanism for generating large-scale spiral arms, crescent-like structures, warped disks, and enhanced accretion activity. The long survival time of streamer-induced kinematic signatures further implies that the effects of streamer accretion may remain observable long after the infall itself has ceased. Consequently, some of the kinematic and morphological perturbations observed in disks without currently detected streamers may represent fossil signatures of past infall events.

Streamer-driven disk evolution may also have important consequences for planet formation and planet--disk interactions. Although large eccentricities can increase planetesimal collision velocities, gas drag and disk self-gravity may help maintain favorable conditions for planet growth. Furthermore, the global eccentricity induced by streamer accretion can alter gap structures and modify the torques acting on embedded planets, potentially changing their migration behavior. Together, these results suggest that environmental accretion should be considered an important ingredient in models of PPD evolution and planet formation.

Finally, we acknowledge several limitations of the present study. We employ a simplified UCM-based streamer model, which allows for robust parameter control but serves only as a first-order approximation of true disk--environment interactions. Our hydrodynamic simulations also neglect magnetic fields, which are known to influence the morphology and kinematics of infalling material. In addition, our thermodynamic model does not fully capture localized shock heating at the stream--disk interface because the adopted cooling timescale is highly efficient. Future work incorporating more realistic streamer prescriptions, magnetic fields, and radiation-hydrodynamic treatments, together with higher numerical resolution, will be necessary to better understand streamer accretion and its role in shaping PPD evolution.

\color{black}

\section{Acknowledgements}
We are grateful to the referee for detailed and 
constructive comments. We also thank Andrés F. Izquierdo for his valuable assistance with using the \textsc{discminer} package.

This work was supported by the National Research Foundation of Korea (NRF) grant funded by the Korea government (MSIT) (grant numbers RS-2024-00416859 and RS-2026-25490557).

\bibliographystyle{aasjournalv7.1}
\bibliography{ms}{}


\appendix
\section{Post-Infall Angular Momentum Deficit} \label{app:amd}

In our simulations, both the disk eccentricity and warp decay after infall ceases. To assess whether this damping is physical or numerical, we calculate the angular momentum deficit (AMD), a quantity frequently used to measure the system's total eccentricity and warp \citep[e.g.,][]{DO2021,Deng2026}. The AMD is defined relative to circular, coplanar orbits with the same semi-major axes as
\begin{equation}
    \text{AMD} \equiv \sum_i m_i\sqrt{GM_*a_i}\left(1-\sqrt{1-e_i^2}\cos i_i\right),
\end{equation}
where $m_i$ is the gas mass in each cell, and $a_i$, $e_i$, $i_i$ are the semi-major axis, eccentricity, and orbital inclination, respectively, calculated from the position and velocity vectors of each cell. For $e_i,i_i\ll1$, the AMD reduces to ${\text AMD} \simeq \frac{1}{2} \sum_i m_i\sqrt{GM_*a_i}
\left(e_i^2+i_i^2\right)$, showing that the AMD provides a measure of the combined eccentricity and inclination of the disk.

\begin{figure}[htb!]
    \epsscale{0.5}
    \plotone{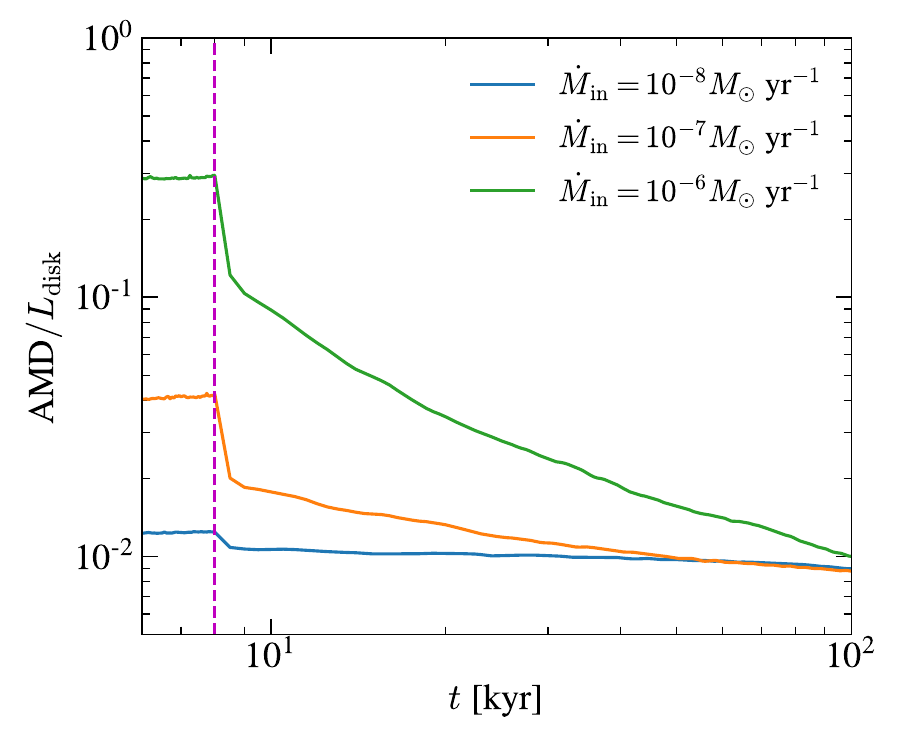}
    \caption{Temporal evolution of AMD normalized by the total disk angular momentum $L_{\mathrm{disk}}$ for models with $\dot{M}_{\mathrm{in}}=10^{-8}\mdotyr$ (\textit{blue}), $10^{-7}\mdotyr$ (\textit{orange}), and $10^{-6}\mdotyr$ (\textit{green}). The vertical dashed line at $t=8\kyr$ marks the end of the infall phase. The normalized AMD declines steadily in all fiducial models after infall ceases.}
    \label{fig:app_AMD}
\end{figure}

\cref{fig:app_AMD} presents the temporal evolution of the AMD normalized by the total disk angular momentum $L_{\mathrm{disk}}$. Because the nearly parabolic accretion streamers supply material with high AMD, the disk AMD at the end of the infall phase ($t=8\kyr$) increases with increasing infall rates. After infall ceases, the AMD initially drops rapidly and then continues to decline steadily through the end of the simulation. Although the disk material lost through the inner radial boundary carries away AMD, the rate of AMD loss does not correlate well with the stellar accretion rate shown in \cref{fig:accretionrate}. This suggests that mass loss through the inner boundary alone cannot account for the post-infall decay.

One possible physical mechanism for the AMD decay is the parametric instability, in which pairs of inertial waves are excited via parametric resonance with oscillatory disk motions, generating small-scale turbulence. Such instability can occur in both eccentric \citep{P2005, BO2014} and warped \citep{GG2000} disks, and the resulting turbulence can damp both eccentricity and warp. Parametric instability in eccentric or warped disks has also been demonstrated in numerical simulations \citep{DO2021, Deng2026}.

Our simulations, however, are unlikely to adequately resolve the parametric instability. \citet{DO2021} noted that previous global grid simulations with fewer than $\sim 10$ cells per scale height failed to capture the instability, whereas local simulations that successfully resolved it employed at least 32 cells per scale height. Our resolution of $\sim7$ cells per scale height at $R_c$ is therefore insufficient to reliably capture this instability.

Numerical diffusion is therefore a likely contributor to the post-infall AMD decay. Grid-scale dissipation can artificially damp disk eccentricity \citep{BO2016, LD2023} and produce a spurious decrease in disk inclination \citep{KD2024}. In addition, the wave-damping zone near the inner radial boundary, which relaxes $\rho$ and $v_{\phi}$ toward their initial profiles, may circularize the inner disk and further reduce the AMD \citep{PF2026}. These numerical effects may cause our simulations to underestimate the persistence of streamer-induced perturbations, which could survive significantly longer in higher-resolution simulations.

\end{document}